\documentclass[%
reprint,
superscriptaddress,
amsmath,amssymb,
aps,
pra,
]{revtex4-2}
\usepackage{xcolor}
\usepackage{array}
\usepackage{amsmath}
\usepackage{dcolumn}

\usepackage{graphicx}% Include figure files
\usepackage{dcolumn}% Align table columns on decimal point
\usepackage{bm}% bold math
\usepackage[breaklinks,colorlinks,bookmarks=false,citecolor=blue,linkcolor=red,urlcolor=blue]{hyperref}
\usepackage{tikz}

\newcommand{\abs}[1]{\left|#1\right|}

\newcommand{\figdir}{images}

\usepackage{color}
\definecolor{Blue}{rgb}{0.3,0.3,0.9}
\definecolor{Red}{rgb}{0.9,0.3,0.3}
\definecolor{Green}{rgb}{0.3,0.6,0.3}

\begin{document}

\title{Phase determination with and without deep learning}

\author{Burak Çivitcioğlu}
\email{Burak.Civitcioglu@cyu.fr}

\affiliation{Laboratoire de Physique Th\'eorique et Mod\'elisation, CNRS UMR 8089, CY Cergy Paris Universit\'e, 95302 Cergy-Pontoise, France}%%

\author{Rudolf A.\ R\"{o}mer}%
\email{R.Roemer@warwick.ac.uk}
\affiliation{
  Department of Physics, University of Warwick, Coventry, CV4 7AL,
United Kingdom
}

\author{Andreas Honecker}
\email{Andreas.Honecker@cyu.fr}
\affiliation{Laboratoire de Physique Th\'eorique et Mod\'elisation, CNRS UMR 8089, CY Cergy Paris Universit\'e, 95302 Cergy-Pontoise, France}%%

\date{March 14, 2024}

\begin{abstract}
Detection of phase transitions is a critical task in
statistical physics, traditionally pursued through analytic methods and direct numerical simulations. 
Recently, machine-learning techniques have emerged as promising tools in this context, with a particular focus on supervised and unsupervised learning methods, along with non-learning approaches. In this work, we study the performance of unsupervised learning in detecting phase transitions in the $J_1$-$J_2$ Ising model on the square lattice. The model is chosen due to its simplicity and complexity, %respectively, 
thus providing an understanding of the application of machine-learning techniques in both straightforward and challenging scenarios.
We propose a simple method based on a direct comparison of configurations. The reconstruction error, defined as the mean-squared
distance
between two configurations, is used to determine the critical temperatures ($T_c$). The results from the comparison of configurations are contrasted with that of the configurations generated by variational autoencoders.  Our findings highlight that for certain systems, a simpler method can yield results comparable to more complex neural networks. This work contributes to the broader understanding of machine-learning applications in 
statistical physics and introduces an efficient approach to the detection of phase transitions using machine determination techniques.
\end{abstract}

%\keywords{Suggested keywords}%Use showkeys class option if keyword display desired
\maketitle

%\tableofcontents

%%%%%%%%%%%%%%%%%%%%%%%%%%%%%%%%%%%%%%%%%%%%%%%%%%%%%%%%%%%%%%%%%%%%%%
\section{Introduction}
\label{sec:introduction}
%%%%%%%%%%%%%%%%%%%%%%%%%%%%%%%%%%%%%%%%%%%%%%%%%%%%%%%%%%%%%%%%%%%%%%

Identification of critical points separating distinct phases of matter is a central pursuit in condensed matter and statistical physics \cite{Ashcroft,diep}. This task requires a thorough understanding of the global behavior of the many-body system because phenomena may emerge that
are very difficult to derive from microscopic rules \cite{Anderson72}.
Traditional analytic methods and numerical simulations have proven effective in understanding these complex systems \cite{Yu2016}, but they often come with limitations, particularly in high-dimensional parameter space \cite{gomez2019review}.  

Machine-learning methods, particularly supervised \cite{alpaydin2020introduction} and unsupervised learning techniques \cite{hinton1999}, have in the last years appeared in physics as a novel strategy to bypassing some of these limitations \cite{Mehta_2019,Carleo2019}. They have been shown to yield promising predictions in identifying critical points or phases in parameter space \cite{Carrasquilla2017,Chang2017,Tanaka_2017,Huembeli_2018, Dong_2018, Canabarro_2019, Shinjo2019}, providing an alternative and potentially more efficient way of exploring complex systems.
By now, the evidence in favor of supervised machine-learning methods' efficacy in identifying different phases of a physical system is compelling \cite{Carrasquilla2017, Chang2017, Tanaka_2017, Huembeli_2018, Dong_2018, Canabarro_2019, Shinjo2019}. 
Unsupervised learning and semi-unsupervised learning have similarly demonstrated the ability to reconstruct the outlines of a system's phase diagram. This is facilitated by several strategies, including but not limited to anomaly detection \cite{Wang2016,Kottmann2020,Alexandrou_2019,dAngelo20,Shiina2020,Corte2021}. The potential to identify structural changes within a system further supports the significance of these techniques in modern scientific exploration \cite{Walker_2018,10.1063/9.0000686}.

Among the various models studied in the context of machine learning and statistical physics, the Ising model on the square lattice has served as a benchmark \cite{Wang2016,Carrasquilla2017,Morningstar2017,Efthymiou19,Walker2020,Goel_20,dAngelo20,Shiina2020,Civitcioglu_2022,basu2023machine,naravane2023semisupervised,pavioni2024minimalist} due to its simplicity and the ready availability of its exact solution \cite{Onsager,McCoyWu+1973}.
Let us also mention related work on multi-layer \cite{Rzadkowski_2020} and Potts models \cite{Fukushima_21, Giataganas_2022, Tirelli2022}, where the latter include the Ising model as the $q=2$ case.
Percolation can be considered as the $q\to1$ limit of the Potts model \cite{RevModPhys.54.235}
and yields
another class of models to which machine-learning techniques have been applied \cite{PhysRevE.99.032142,Yu2020,PhysRevE.103.052140,Bayo_2022,Bayo_2023,patwardhan2023machine}.

The $J_1$-$J_2$ Ising model incorporates competing interactions across the diagonals of the squares and presents a more challenging case than the aforementioned ones.
Investigations of this model have a long history in statistical physics
\cite{Swendsen79,PhysRevB.21.1285,PhysRevB.21.1941,PhysRevB.31.5946,PhysRevB.48.3519,Zandvliet06,Malakis2006,PhysRevE.76.021123,DOSANJOS20081180,Kalz_2008,Zandvliet09,PhysRevB.84.174407,PhysRevLett.108.045702,PhysRevB.86.134410,jin2012,PhysRevE.91.032145,RAMAZANOV201635,PhysRevB.101.014453,PhysRevE.104.024118,PhysRevB.104.144429,Zandvliet23,PhysRevE.107.034124,Watanabe23,yoshiyama2023higherorder,gangat2023weak,abalmasov2023free,PhysRevB.109.064422}.
It was observed early on \cite{Swendsen79,PhysRevB.21.1285} that, with $J_1$ denoting the nearest-neighbor interaction, the competing second-neighbor interaction $J_2$ gradually suppresses the ordering temperature, until it vanishes completely when $J_2=\abs{J_1}/2$. Furthermore, beyond this point, a new ordered phase called the ``superantiferromagnetic phase'' appears.
The universality class of the transition into the superantiferromagnetic phase has been investigated early on  \cite{Swendsen79,PhysRevB.31.5946}, but continues to attract attention
\cite{PhysRevB.48.3519,Malakis2006,PhysRevE.76.021123,DOSANJOS20081180,Kalz_2008,PhysRevB.84.174407,PhysRevLett.108.045702,PhysRevB.86.134410,jin2012,PhysRevE.91.032145,RAMAZANOV201635,PhysRevB.101.014453,PhysRevE.104.024118,Watanabe23,yoshiyama2023higherorder,gangat2023weak} since its nature remains controversial.
There is at least also one investigation of this model on the D-wave quantum annealer \cite{Park22}
and a small number of machine-learning investigations
\cite{Corte2021,basu2023machine}.

In this work, we focus on the square-lattice $J_1$-$J_2$ Ising model, using machine-learning techniques to predict phase transitions and construct the phase diagram.
We adopt the approach of detecting criticality based on the reconstruction error ($\mathcal{E}$), defined as the mean-squared
distance
between two spin configurations, by comparatively using two machine determination methods \cite{Acevedo_2021_2}.
The first method is the Variational Autoencoder (VAE), a type of neural network that reconstructs a given predicted state after being trained on a selected set of states \cite{Kingma2014}.
We use the TensorFlow interface to implement our VAE \cite{tensorflow2015-whitepaper}.
The second machine-determination method is simpler, based on using a configuration comparison (CMP) and does not require training nor any bespoke machine-learning tools.

Our study aims to provide insights into the capabilities of these computational methods in the context of phase transition detection and
we conclude that the simpler method (CMP) can achieve success rates that can be compared to that of the VAEs.

%%%%%%%%%%%%%%%%%%%%%%%%%%%%%%%%%%%%%%%%%%%%%%%%%%%%%%%%%%%%%%%%%%%%%%
\section{The $J_1$-$J_2$ Ising Model}
\label{sec:model}
%%%%%%%%%%%%%%%%%%%%%%%%%%%%%%%%%%%%%%%%%%%%%%%%%%%%%%%%%%%%%%%%%%%%%%t

%%%%%%%%%%%%%%%%%%%%%%%%%%%%%%%%%%%%%%%%%%%%%%%%%%%%%%%%%%%%%%%%%%%%%%
\begin{figure}
\begin{center}
\includegraphics[width=0.6\columnwidth]{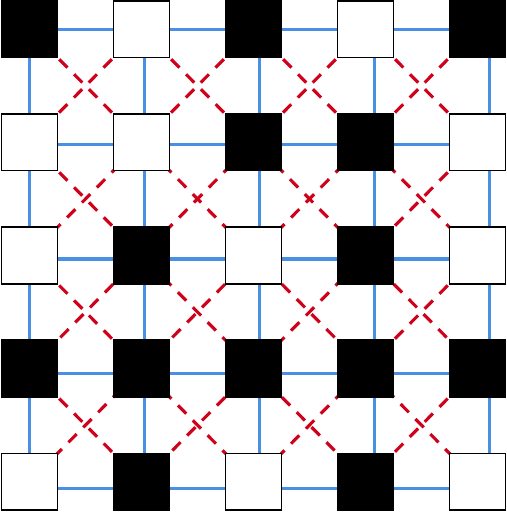}
\caption{Schematic representation of the $J_1$-$J_2$ Ising model on a $5\times 5$ square lattice. The squares designate the classical spins with
black and white corresponding to up and down states, respectively, chosen to illustrate one spin configuration.
The solid blue and dashed red lines denote the interactions of nearest neighbors $J_1$ and next-nearest neighbors $J_2$, respectively.
}
\label{fig:schematic-j1j2}
\end{center}
\end{figure}
%%%%%%%%%%%%%%%%%%%%%%%%%%%%%%%%%%%%%%%%%%%%%%%%%%%%%%%%%%%%%%%%%%%%%%

The $J_1$-$J_2$ Ising model serves as a simple but non-trivial system to illustrate phase transitions, especially those associated with magnetic behavior.
As presented in Fig.~\ref{fig:schematic-j1j2}, it adds the complexity of second-nearest-neighbor interactions to the traditional nearest-neighbor Ising model.
The Hamiltonian of the $J_1$-$J_2$ Ising model is expressed as 
\begin{equation}
H_{J_1J_2} = -J_1 \,\sum_{\langle i,j \rangle} s_i\, s_j + J_2 \,\sum_{\langle\langle i,j \rangle\rangle} s_i \,s_j\, .
\label{eq:hamiltonian}
\end{equation}
Here, $s_i$ represents the spin at site $i$, which can be either up ($+1$) or down ($-1)$; $\langle i,j \rangle$ refers to nearest-neighbor pairs, $\langle\langle i,j \rangle\rangle$ denotes next-nearest neighbor pairs, while $J_1$, $J_2 \geq 0$  signify the interaction strengths between the nearest and next-nearest neighbors, respectively.
The sign convention of the Hamiltonian in Eq.~(\ref{eq:hamiltonian}) leads to a ferromagnetic coupling for $J_1$ pairs while next-nearest neighbors prefer to align in an antiferromagnetic structure.
We investigate square lattices of size $L \times L$ (linear extent $L$) and impose periodic boundary conditions.

The $J_1$-$J_2$ Ising model reduces to the canonical Ising model when considering only nearest-neighbor interactions.
In other words, the traditional Ising model can be regarded as the $J_2=0$ special case of the $J_1$-$J_2$ Ising model.
We show in appendix \ref{app:A} that the case of antiferromagnetic nearest-neighbor coupling $J_1<0$ can be mapped to the case of ferromagnetic
coupling $J_1 > 0$. Consequently, the phase diagram is independent of the sign of $J_1$ although the exact nature of each phase depends on the sign conventions. In the present
work, we will focus on the ferromagnetic case $J_1>0$, and  present results, in particular for the phase diagram, in units of $\left|J_1\right|=1$ in order to emphasize that equivalent results would be obtained in the case of antiferromagnetic $J_1<0$.

%%%%%%%%%%%%%%%%%%%%%%%%%%%%%%%%%%%%%%%%%%%%%%%%%%%%%%%%%%%%%%%%%%%%%%
\begin{figure}[tb]
     \begin{center}
    \includegraphics[width=1\columnwidth]{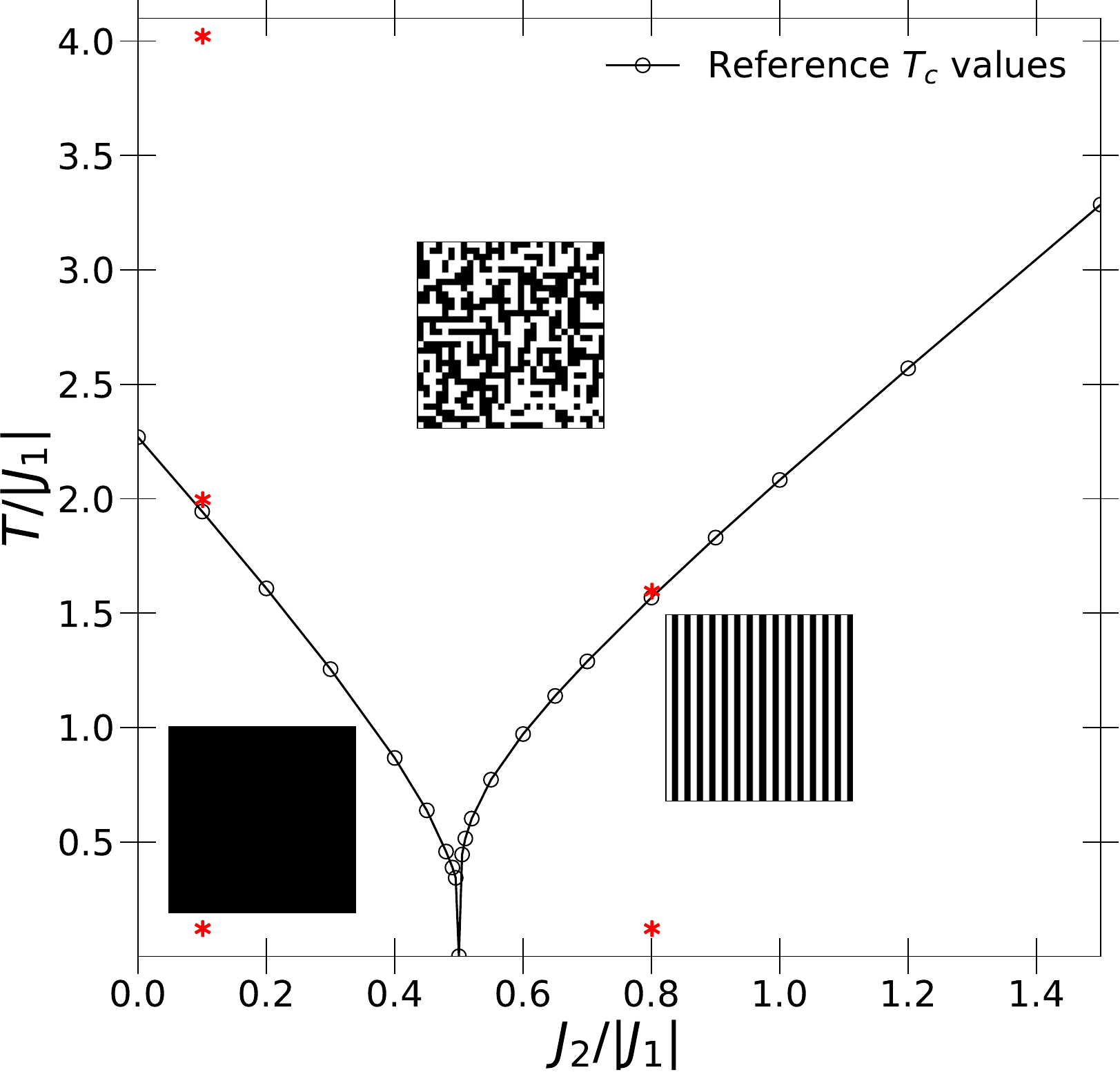}
    \end{center}
  \caption{Phase diagram of the $J_1$-$J_2$ Ising model on the periodic square lattice,
  with sample configurations from each of the three phases shown.
  The top \emph{paramagnetic} configuration was obtained at $J_1 = 1$, $J_2 = 0.1$, and $T=4>T_c$.
  The bottom left \emph{ferromagnetic} configuration corresponds to $J_1 = 1$, $J_2 = 0.1$, and $T=0.1 < T_c$.
  The bottom right configuration was found for $J_1 = 1$, $J_2 = 0.8$, and $T=0.1 < T_c$. 
  This configuration illustrates the \emph{superantiferromagnetic} phase.
  The reference $T_c$ data is based on Ref.~\cite{Kalz_2008} and shown by circles connected by lines as guide for the eye. 
  The five red stars ($*$) locate the $(T,J_2)$ positions for selected configurations further detailed in Fig.\ \ref{fig:sample-configurations}.
  }
    \label{fig:phase-diagram}
\end{figure}
%%%%%%%%%%%%%%%%%%%%%%%%%%%%%%%%%%%%%%%%%%%%%%%%%%%%%%%%%%%%%%%%%%%%%%

%%%%%%%%%%%%%%%%%%%%%%%%%%%%%%%%%%%%%%%%%%%%%%%%%%%%%%%%%%%%%%%%%%%%%%
\begin{figure*}[tb!]
     \begin{center}
    \raisebox{0\columnwidth}{(a) }\includegraphics[width=0.25\columnwidth]{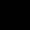}\hfil%
    \raisebox{0\columnwidth}{(b) }\includegraphics[width=0.25\columnwidth]{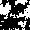}\hfil%
    \raisebox{0\columnwidth}{(c) }\includegraphics[width=0.25\columnwidth]{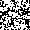}\hfil%
    \raisebox{0\columnwidth}{(d) }\includegraphics[width=0.25\columnwidth]{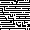}\hfil%   
    \raisebox{0\columnwidth}{(e) }\includegraphics[width=0.25\columnwidth]{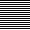}
    \end{center}
    \begin{center}
    \raisebox{0\columnwidth}{(f) }\includegraphics[width=0.25\columnwidth]{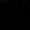}\hfil%
    \raisebox{0\columnwidth}{(g) }\includegraphics[width=0.25\columnwidth]{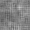}\hfil%
    \raisebox{0\columnwidth}{(h) }\includegraphics[width=0.25\columnwidth]{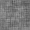}\hfil%
    \raisebox{0\columnwidth}{(i) }\includegraphics[width=0.25\columnwidth]{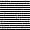}\hfil%   
    \raisebox{0\columnwidth}{(k) }\includegraphics[width=0.25\columnwidth]{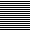}
    \end{center}
    \begin{center}
    \end{center}
  \caption{(a-e) Five illustrative spin configurations of the $J_1$-$J_2$ Ising model on a periodic $30 \times 30$ square lattice for $(T,J_2)$ values as indicated in Fig.\ \ref{fig:phase-diagram}.
  (f-k) Predicted spin configurations from (a-e), respectively, using the single VAE learning described in section \ref{sec:singleVAE}.
  Panels (a, b, c) keep $J_2 = 0.1$ fixed while increasing $T$ from 
  (a) ferromagnetic at $T=0.1$ to 
  (b) $T=1.975$ near the ferromagnetic-to-paramagnetic transition and 
  (c) a configuration deep in the paramagnetic phase at $T=4.0$.
  Panels (d, e) have $J_2=0.8$ and then decrease $T$ from 
  (d) $T=1.575$ near the paramagnetic-to-superantiferromagnetic transition to
  (e) a superantiferromagnetic configuration at $T=0.1$. 
  In (f-k), the parameters are as in (a-e).
  In all cases, the black squares correspond to up spins while white is for down spins as in Fig.\ \ref{fig:schematic-j1j2}. In (f-k), the values $]-1, 1[$ correspond to the gray scale in the panels. }
    \label{fig:sample-configurations}
\end{figure*}

%%%%%%%%%%%%%%%%%%%%%%%%%%%%%%%%%%%%%%%%%%%%%%%%%%%%%%%%%%%%%%%%%%%%%%
\subsection{Phase diagram}
\label{sec:model-phasediagram}

Figure \ref{fig:phase-diagram} recalls the well-studied phase diagram of the $J_1$-$J_2$ Ising model. Here, we use the previously computed high-precision transition temperatures $T_c$ from Ref.~\cite{Kalz_2008}
for reference. We should note that Ref.~\cite{Kalz_2008} uses different sign conventions for $J_1$ from the present work, resulting in different terminology, but we recall that the two cases are equivalent.
Some more recent numerical investigations such as Refs.~\cite{PhysRevB.84.174407,PhysRevLett.108.045702,jin2012,PhysRevE.104.024118,yoshiyama2023higherorder} may provide more accurate estimates of the transition temperatures, but any potential differences are so small that they are irrelevant for the present purposes.

The $J_1$-$J_2$ Ising model exhibits three distinct phases that we illustrate, by one representative spin configuration each, in Fig.~\ref{fig:phase-diagram}.
We have generated these by the Monte Carlo method described in the next section.
The paramagnetic phase appears at sufficiently high temperatures, namely $T>T_c$ irrespective of the values of the interaction constants. This phase is characterized by the absence of long-range order.
The next phase is the ferromagnetic one, characterized by a preference of the neighboring spins to align. In the $T=0$ ground state, spins are
perfectly aligned, yielding an energy per site of
\begin{equation}
e_{\rm ferro} = -2\,J_1 + 2\,J_2 \, .
\label{eq:Esuper}
\end{equation}
The finite-temperature phase transition to this ferromagnetic phase starts at the exactly known value $T_{c,{\rm Ising}}/\abs{J_1} = 2/{\ln(1+\sqrt{2})} \approx 2.269$ for $J_2=0$ \cite{Onsager} and is gradually suppressed by a competing $J_2>0$.
The third and last phase is known as the superantiferromagnetic phase \cite{PhysRevB.21.1285}. Here, the $J_1$ and $J_2$ interactions compete: $J_1$ prefers to align nearest-neighbor spins
parallel while $J_2$ tries to enforce an antiparallel alignment of next-nearest neighbor spins. In the superantiferromagnetic state,
either vertical or horizontal stripes composed of opposing spins are formed, thus satisfying all $J_2$ interactions and half of the $J_1$ interactions.
At $T=0$, this order is perfect, yielding an energy per site of
\begin{equation}
e_{\rm super} = - 2\,J_2 \, .
\label{eq:Eferro}
\end{equation}
Further examples of spin configurations in these three phases are given in Fig.~\ref{fig:sample-configurations} with a more detailed discussion to follow later in this section.

When $J_2=\left|J_1\right|/2$ for $T=0$, we see from Eqs.~\eqref{eq:Esuper} and \eqref{eq:Eferro} that the ground-state energies become degenerate, corresponding to the transition point between ferromagnetic and superantiferromagnetic phases.
Numerical investigations \cite{Kalz_2008,PhysRevE.104.024118,yoshiyama2023higherorder,PhysRevB.109.064422}
indicate that there is no finite-temperature phase transition exactly at $J_2=\left|J_1\right|/2$
and that the critical temperature $T_c$ is suppressed to $T_c=0$ when approaching $J_2=\left|J_1\right|/2$ from either ordered phase.

We emphasize again that both signs of $J_1$ yield equivalent physics, in particular an equivalent phase diagram. We therefore focus on the case $J_1>0$.
By contrast, the sign of $J_2$ matters, and this needs to compete with the nearest-neighbor coupling $J_1$ in order to give rise to a new phase. With the conventions used in Eq.~(\ref{eq:hamiltonian}), this corresponds to the case $J_2>0$. A ferromagnetic $J_2<0$ would just stabilize the ordered state of the simple nearest-neighbor Ising model on the square lattice, but yield no new physics. For this reason, we also focus on the case $J_2 \ge 0$.

Figure \ref{fig:sample-configurations} shows further examples of configurations, emphasizing changes upon approaching the phase transition.
Panels (a) and (e) show configurations at low temperatures in the ferromagnetic and superantiferromagnetic phase, respectively.
These are similar to the configurations already shown in Fig.~\ref{fig:phase-diagram}, except that in the superantiferromagnetic case the stripes in the examples are rotated by $90^\circ$.
Panels (b) and (d) of Fig.~\ref{fig:sample-configurations} show configurations at higher temperatures, closer to the critical temperature $T_c$.
Here one observes fluctuations on top of the ordered background.
Finally, Fig.~\ref{fig:sample-configurations}(c) shows another example for a configuration in the high-temperature paramagnetic phase.

%%%%%%%%%%%%%%%%%%%%%%%%%%%%%%%%%%%%%%%%%%%%%%%%%%%%%%%%%%%%%%%%%%%%%%
\subsection{Monte Carlo Method}
\label{sec:model-mcmethod}

To generate the necessary input data for the machine determination models, we utilize the Metropolis algorithm, a well-established method in the realm of computational physics for simulating thermal systems~\cite{Metropolis_1953,Newman99,Berg_2004,Landau_2014}.

In the present investigation, we initially focus on a system size of $30\times30$ with periodic boundary conditions.
This choice is motivated by the machine-learning frameworks being tailored for images of similar size.
Indeed, previous related work \cite{Corte2021,Acevedo_2021_2} for the $J_1$-$J_2$ Ising model on the square and honeycomb lattices also employed $L=30$.
In order to assess the influence of the size of the system, we also investigate  $60\times60$ and $120\times120$ square lattices, again with periodic boundary conditions.

Equilibration of the simulations can be difficult, in particular in the regime of
$J_2 \approx \left|J_1\right|/2$ \cite{Kalz_2008}.
In order to ensure proper thermalization within a simple single-spin flip Metropolis scheme,
we proceed as follows: we fix $J_2/\left|J_1\right|$\ and start from a high initial temperature $T/\left|J_1\right|=100$, where the spin
configurations are essentially random.
Next, we gradually lower the temperature to $T/\left|J_1\right|=4$ over the course of 1000 Monte Carlo sweeps (MC sweeps, i.e., complete $L\times L$ spin updates). 
Then we start the data collection phase:
for each $T$, we first thermalize for another $3000$ MC sweeps. Next, we collect five spin configurations at a given $T$, spaced by $1000$ MC sweeps between each measurement to ensure the statistical independence of the configurations.
Finally, temperature is lowered by $\Delta T/\left|J_1\right| = 0.025$ and the procedure repeated,
until we reach $T/\left|J_1\right|=0.1$.
This yields a set ${\cal T}$ of $|\mathcal{T}|=157$ temperatures with
$T/\left|J_1\right| \in [0.1,4]$ for $T \in {\cal T}$.
The procedure is repeated with different random numbers until we have $C=40$ configurations for each temperature at the given value of $J_2/\left|J_1\right|$.

We then select another $J_2/\left|J_1\right| \in {\cal J}_2$ and collect spin configurations as above. Here, ${\cal J}_2$ denotes the set  ${\cal J}_2 = \{
0$, $0.1$, $0.2$, $0.3$, $0.4$, $0.45$, $0.48$, $0.49$, $0.495$, $0.5$, $0.505$,
$0.51$, $0.52$, $0.55$, $0.6$, $0.65$, $0.7$, $0.8$, $0.9$, $1$, $1.2$, $1.5 
\}$ with $|\mathcal{J}_2|=22$ distinct values.
In total, this results in a dataset containing $|\mathcal{T}|\times |\mathcal{J}_2| \times C = 157 \times 22 \times 40 = 138\,160$ independent configurations for a given system size. We shall denote this \emph{global} dataset as $\rho_{{G}}$.
The examples shown in Figs.~\ref{fig:phase-diagram} and \ref{fig:sample-configurations} were taken from $\rho_{\text{G}}$.
We note that the configurations are stored in an exact numeric form, and not as potentially lossy images, as the machine-learning context might suggest \cite{Li_2020}.

%%%%%%%%%%%%%%%%%%%%%%%%%%%%%%%%%%%%%%%%%%%%%%%%%%%%%%%%%%%%%%%%%%%%%%
\section{Machine-learning approaches}
\label{sec:machine-learning}
%%%%%%%%%%%%%%%%%%%%%%%%%%%%%%%%%%%%%%%%%%%%%%%%%%%%%%%%%%%%%%%%%%%%%%

\begin{figure}[t!]
\begin{center}
\includegraphics[width=\columnwidth]{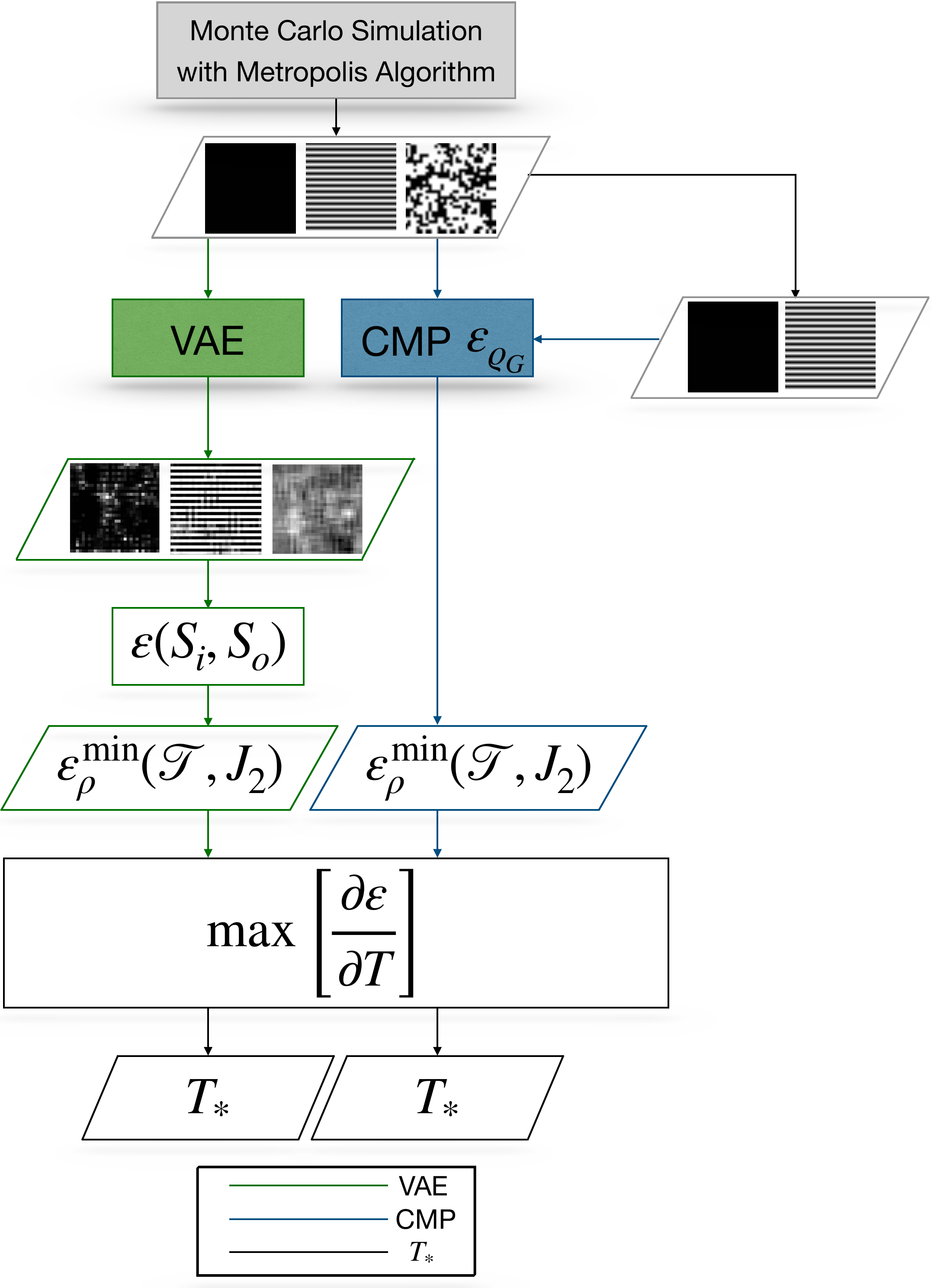}
\caption{Workflows of the two machine-``learning'' approaches implemented in the present work:
We begin with an input dataset that is processed in the first approach by a VAE to generate a reconstructed set of configurations. In the second approach, we compare the input images against some reference configurations (CMP). For both pathways, we calculate the reconstruction error ${\cal E}$ for each configuration, resulting in a set of reconstruction error distributions. By analyzing the derivative of the distribution, we finally pinpoint the temperature with the largest derivative as the predicted critical temperature $T_c$ for a given value of $J_2/\left|J_1\right|$.
}
\label{fig:schematics-flow-diagram}
\end{center}
\end{figure}

The datasets described in the preceding section will form the basis for two independent machine-determination approaches to the phase diagram of the $J_1$-$J_2$ Ising model.
In principle, we would like to explore fully unsupervised approaches, but we note that some prior knowledge about the phase diagram did go into the generation of the underlying dataset, namely the relevant range of temperatures $T$ and the required resolution of coupling ratios $J_2/\left|J_1\right|$.
Indeed, the workflow presented in Fig.~\ref{fig:schematics-flow-diagram} shows that generation of the input data is the beginning of the study. 
%%%%%%%%%%%%%%%%%%%%%%%%%%%%%%%%%%%%%%%%%%%%%%%%%%%%%%%%%%%%%%%%%%%%%%
%%%%%%%%%%%%%%%%%%%%%%%%%%%%%%%%%%%%%%%%%%%%%%%%%%%%%%%%%%%%%%%%%%%%%%

%%%%%%%%%%%%%%%%%%%%%%%%%%%%%%%%%%%%%%%%%%%%%%%%%%%%%%%%%%%%%%%%%%%%%%
\begin{figure*}[tb!]
\begin{center}
\includegraphics[width=2\columnwidth,trim=0 90 0 75,clip]{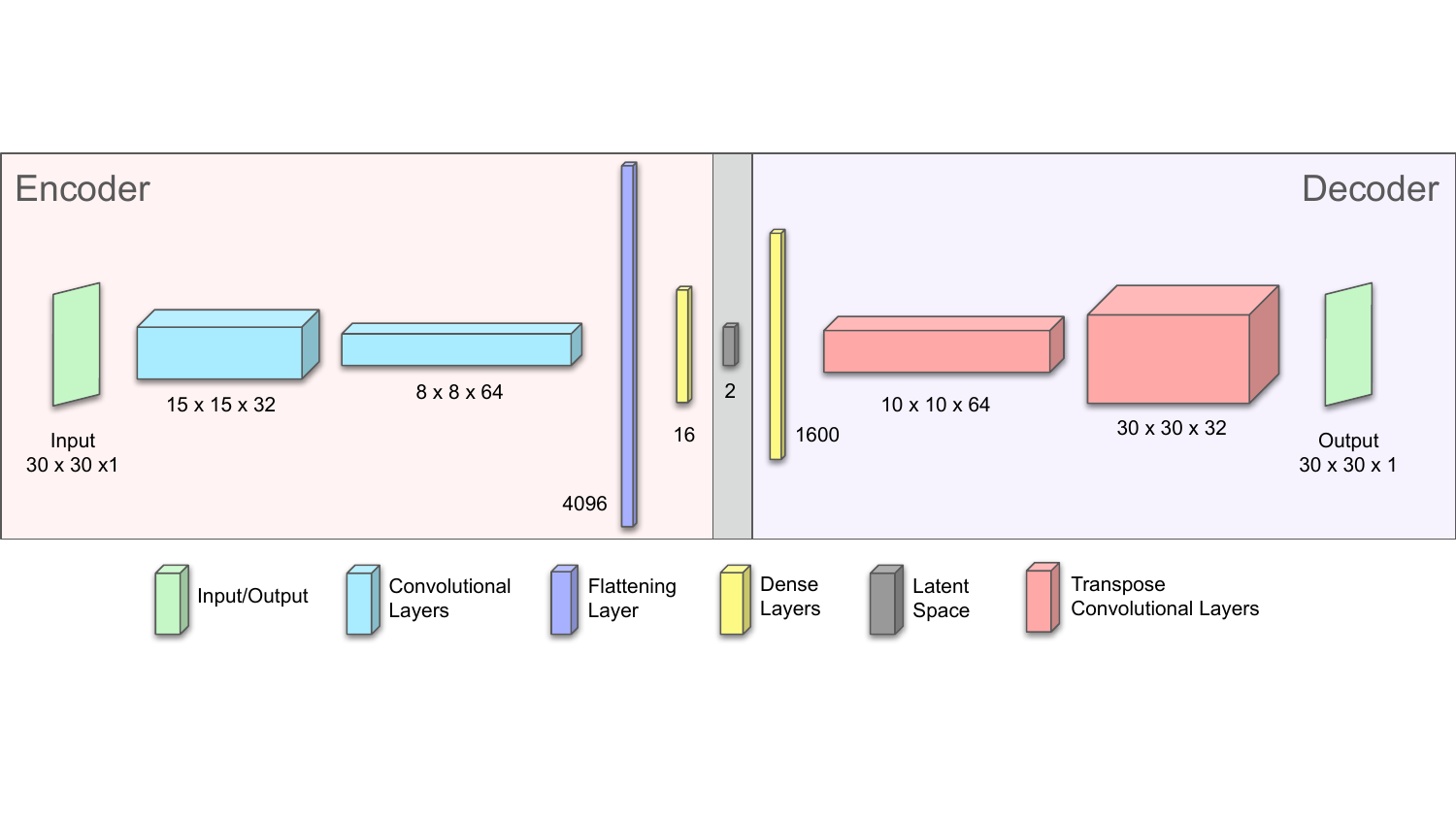}
\end{center}
  \caption{
Network architecture of the variational autoencoder for the $30 \times 30$ lattice.
The encoder consists of two convolutional layers with 32 and 64 filters, ReLu activation function, and padding, followed by a flattening layer and a dense layer with 16 units. The dense layer outputs the mean and log-variance parameters to the latent space, from which the latent vector $\mathbf{z}$ is sampled. The decoder, starting with a dense layer and a reshape layer, uses transposed convolutional layers to reconstruct the input image from $\mathbf{z}$. The final layer applies a sigmoid activation function to produce output values between 0 and 1, which are then rescaled to the interval $[-1,1]$.
  }
    \label{fig:schematics-VAE}
\end{figure*}
%%%%%%%%%%%%%%%%%%%%%%%%%%%%%%%%%%%%%%%%%%%%%%%%%%%%%%%%%%%%%%%%%%%%%%

We will employ two distinct computational methods:
the first approach follows Refs.~\cite{Corte2021,Acevedo_2021_2} and uses a deep-learning model, viz.\ a variational autoencoder (VAE)
that produces a predicted output spin configuration for each given input configuration.
The second approach is much simpler, namely just a direct comparison of configurations, but has to the best of our knowledge not been implemented previously.
The generated datasets will be used, as shown in Fig.~\ref{fig:schematics-flow-diagram} (to be discussed in more detail later in Sec.~\ref{sec:singleVAE-global}), as the training data for the VAE and the reference configurations for the simpler proposed method, direct comparison of configurations. Consequently, the reconstruction errors are computed, from which the phases can be identified. Then as the final step in the workflow we
estimate the critical temperatures for each $J_2$ value.

%%%%%%%%%%%%%%%%%%%%%%%%%%%%%%%%%%%%%%%%%%%%%%%%%%%%%%%%%%%%%%%%%%%%%%
\subsection{Variational Autoencoder}
\label{sec:ml-vae}

A VAE is a relatively recent deep-learning architecture that combines standard compression techniques with the regularization strategies of machine learning, serving also as a generative model \cite{Kingma2014,chou2019generated}. In brief, it consists of an \emph{encoding} multilayered neural network that, upon training with input data, produces output parameters for a variational distribution. These parameters characterize a low-dimensional probabilistic distribution space, known as \emph{latent space}.
The \emph{decoding} part of the VAE then is again a deep neural network architecture that generates the reconstructed output data from the latent space, taking samples from the latent space rather than deterministic points. 
Clearly, when the dimension $d$ of the latent space is much smaller than the information content of the input data, this procedure will lead to some information loss. Hence, one is trying to construct en- and decoders such that upon encoding a maximum of information is kept while upon decoding a minimum of error is introduced into the output data.
In order to optimally train the model to generate such a VAE, two loss measures are employed. The \emph{reconstruction error} $\varepsilon$ (see Sec.~\ref{sec:ml-re} for details)
quantifies the difference between input and output data upon training. In addition, the so-called Kullback-Leibler divergence \cite{kullback}, which acts as a regularization term,  assures the regularization of the latent space, making it approximate standard normal distributions \cite{Kingma2014}. In practice, during training one minimizes a total loss $\ell$ that combines the reconstruction loss $\ell_{\varepsilon}$ in the final layer from $\varepsilon$ as well a Kullback-Leibler loss $\ell_\text{KL}$ \cite{Kingma2014}, such that $\ell=  \ell_{\varepsilon} + c \, \ell_\text{KL}$, where $c$ a constant. We have tried different values of $c\in \{1 \dots 10\}$ and observed that it does not yield any notable changes in the results, therefore we fix $c=1$. 

In the present case, our input data consists of spin configurations with dimensions $(L, L, 1)$ with a simple $\pm 1$ binary value for each of the $L^2$ spins. This is similar to standard black-and-white images, in particular after normal batch-normalization operations in the encoding \cite{luhman2023high}. Hence, we are using in the following the same regularization strategy as used in VAEs for image reconstructions (\textsc{ImageNet} \cite{deng2009imagenet}), employing Gaussian distributions in latent space.  We have also checked that a latent space dimension up to $d=8$ reproduces similar results, but we find that $d=2$, as shown in Fig.~\ref{fig:schematics-VAE}, is sufficient and often better in distinguishing
phases. We suggest that this is due to the relatively small number of phases for the $J_1$-$J_2$ model. We note that $d=2$ has also been used with good accuracy in the MNIST ten-numerals-recognition challenge \cite{Asperti2021,Davidson2018HypersphericalVA}.

Figure~\ref{fig:schematics-VAE}
shows the architecture of the VAE that we have used in our study of the $30 \times 30$ lattice.
For the other system sizes we have used structurally the same network architecture with the parameters adjusted for the different input sizes.
In order to keep the parameters padding, kernel size and stride
for the input size $L=120$
the same as $L=30$ and $L=60$, we added a third convolutional layer. The network architecture for $L=30$ begins with two convolutional layers with 32 and 64 filters, respectively, each with a $3\times3$ kernel size and a stride of 2. Both layers apply a rectifier linear unit or ReLu activation function that is defined as $\mathrm{ReLU}(x) = \max(0,x)$ and use padding to preserve spatial dimensions. After these convolutional layers, a flattening layer reshapes the 3D output into a 1D tensor, which then feeds into a dense layer with 16 units and again $\mathrm{ReLU}$ activation. This layer prepares the data for conversion into the latent space. The final two layers of the encoder output two sets of parameters: the mean, $\mathbf{z}_{\text{mean}}$, and the log-variance, $\mathbf{z}_{\text{log var}}$, of the latent space. These parameters define a sum of Gaussian distributions, from which we sample using the reparameterization, generating the latent vector $\mathbf{z}$ \cite{Kingma2014}.
The decoder then aims to reconstruct the input spin configuration (image) from the compressed information in the latent space.
The architecture of the decoder that we used in our study starts with a dense layer that up-samples $\mathbf{z}$ to dimensions of $5 \times 5 \times 64$. A reshape layer follows, converting the 1D tensor back into a 3D tensor. After this, two sets of transposed convolutional layers with 64 and 32 filters, respectively, are applied, each with a $3\times3$ kernel size and $\mathrm{ReLU}$ activation. The first has a stride of 2 and the second has a stride of 3, and both utilize padding. The final layer of the decoder uses another transposed convolutional layer with one filter to generate the output image, applying a sigmoid activation function defined as
$ \sigma(x) =  {\mathrm{1} }/({\mathrm{1} + \mathrm{e}^{-x} })  $ to ensure that the output values fall within the range between 0 and 1. 

%%%%%%%%%%%%%%%%%%%%%%%%%%%%%%%%%%%%%%%%%%%%%%%%%%%%%%%%%%%%%%%%%%%%%%
\subsection{Reconstruction error}
\label{sec:ml-re}

The typical reconstruction error used in many image-based VAE applications is known as ``mean-squared error'' (MSE) and is defined as
\begin{equation}
\varepsilon(\mathbf{S}_{\text{i}},\mathbf{S}_{\text{o}}) = \frac{1}{4 L^2} \sum_{l=1}^{L^2}({s}_{\text{i},l}-{s}_{\text{o},l})^2 \, ,
\label{eq:mse}
\end{equation}
where $\mathbf{S}_{\text{i}}=\{{s}_{\text{i},l} \}$ and $\mathbf{S}_{\text{o}}=\{{s}_{\text{o},l} \}$ correspond to the input and output configurations of the VAE, respectively.
For two identical spin configurations, $\mathbf{S}_\text{o}=\mathbf{S}_\text{i}$, we obviously have $\varepsilon=0$. For two opposite configurations, $\mathbf{S}_\text{o}=-\mathbf{S}_\text{i}$, we find $\varepsilon=1$ while for two configurations with half the spins identical and half opposite, we have $\varepsilon=0.5$. The latter value is also true when comparing two independent and identically distributed, i.e., random, spin configurations, at least when $L \rightarrow\infty$. We note that in \eqref{eq:mse}, the factor $4$ in the denominator assures that the $\varepsilon$ are normalized similar to the standard results for MSEs where usually comparisons are for values ranging from $0$ to $1$, whereas in the present case the range is $[-1,1]$.
While the spin configurations computed for the $J_1$-$J_2$ model are restricted to values $\pm 1$, no such restriction is in place for the output generated by the VAE and the possible range of $s_{\text{o},l}$ is $[-1, 1]$. Therefore, in principle all values $\in [0, 1]$ are possible when computing $\varepsilon$\ between a spin configuration $\mathbf{S}_\text{i}$ computed from the $J_1$-$J_2$ model, and  $\mathbf{S}_\text{o}$, reconstructed via the VAE. In particular, if the VAE would produce a completely featureless configuration $s_{\text{o},l}=0$ 
for all $l$
in the $L\times L$ lattice, then we would have $\varepsilon=0.25$.

Often, we shall be interested in reconstruction errors originating from differently trained VAEs. For example, we might be interested in selecting a certain smaller region $\mathcal{\rho}$ of the $(\mathcal{T},\mathcal{J}_2)$ parameter space as the space from where the input spin configurations $\mathbf{S}_\text{i}$, used in the training of a VAE, originate. We shall then use $\varepsilon_\rho$ to denote that particular training. 
Furthermore, when testing a particular VAE, we will do so at specific $T$ and $J_2$ values. We note that while we use the term \emph{testing} in the technical ML sense, its physics use is in producing reconstruction errors $\varepsilon_\rho(T,J_2)$ to allow phase reconstruction.
In principle, there is one such $\varepsilon^{(c)}_\rho(T,J_2)$ value for each spin configuration $c$ of the $C=40$ configurations constructed at each point $(T,J_2)$ as discussed in section \ref{sec:model-mcmethod}. Hence it is at this point that a statistical analysis can be applied, e.g., construct mean and minimal estimates, i.e., $\langle \varepsilon_\rho \rangle (T,J_2) = \sum_{c=1}^{C} \varepsilon^{(c)}_\rho(T,J_2)/C$ and $\text{min} (\varepsilon_\rho) (T,J_2) = \text{min} \{ \varepsilon^{(c)}_\rho(T,J_2) \mid
c\in C\}$.

Sets of reconstruction errors shall be denoted $\mathcal{E}$, with 
\begin{equation}
\mathcal{E}_{\rho}(J_2) = \big\{ \{T,\varepsilon_{\rho}(T, J_2)\} \mid 
T \in \mathcal{T} \big\}%.
\label{eq:REsetJ2}
\end{equation}
denoting the set of all reconstruction errors at constant $J_2$ computed for a VAE trained on region $\rho$. Here, when suppressing the explicit mention of the configuration label $c$, we then mean that all $C$ configurations are elements of such a set.
Further, we shall denote by $\mathcal{E}_{\rho}(\mathcal{T},\mathcal{J}_2)$ the set of all $\mathcal{E}_{\rho}(J_2)$ with $J_2 \in \mathcal{J}_2$.
Finally, we define
\begin{eqnarray}
   \mathcal{E}^{\text{min}}_{\rho}(\mathcal{T}, J_2) &=
    &\big\{
    \min_{c \in C}\left[ \varepsilon_\rho (T,J_2) \right] \mid 
    T \in \mathcal{T}
    \big\} , \label{eq:REsetminJ2}\\
    \mathcal{E}^{\text{min}}_{\rho}(\mathcal{T}, \mathcal{J}_2) &=
    &\big\{
    \min_{c \in C}\left[ \varepsilon_{\rho}(T, J_2)\right] \mid
    T \in \mathcal{T}, J_2 \in \mathcal{J}_2
    \big\} \label{eq:REsetminG} .
\end{eqnarray}
Average
$\langle\cdot\rangle$ or maximal $\max[\cdot]$ values can be defined in the same way as above for the minimum.

%%%%%%%%%%%%%%%%%%%%%%%%%%%%%%%%%%%%%%%%%%%%%%%%%%%%%%%%%%%%%%%%%%%%%%
\subsection{Comparing individual spin configurations}
\label{sec:ml-ic}

When training the VAE, the information about the spin configurations in the region $\rho$ is learned, i.e., non-linearly encoded, in the set of parameters of the en-/decoding neural networks and latent space of the VAE. Then, when given an arbitrary input spin configuration at $(T, J_2)$, the VAE will generate a new spin configuration according to the information imprinted on its parameters based on $\rho$, aiming to minimize $\varepsilon_{\rho}(T, J_2)$.
Alternatively, one can also
just use each spin configuration of $\rho$ %into a code 
and then simply compute $\varepsilon_{\rho}(T,J)$ between a test configuration at $(T,J_2)$ and each reference spin configuration. This leads to the set $\mathcal{E}_{\rho}(T,J_2)$ of reconstruction errors. As in the case of the VAE, we can proceed to again define the reconstruction error sets analogously to \eqref{eq:REsetJ2}, \eqref{eq:REsetminJ2}, and \eqref{eq:REsetminG}.

Without further optimization, such a direct comparison of individual spin configurations will scale with the number $|\rho|$ of configurations in $\rho$ and each computation of the reconstruction error in \eqref{eq:mse} uses $O(L^2)$ operations.

%%%%%%%%%%%%%%%%%%%%%%%%%%%%%%%%%%%%%%%%%%%%%%%%%%%%%%%%%%%%%%%%%%%%%%
\section{Reconstruction of the phase diagram using single-region VAEs}
\label{sec:singleVAE}
%%%%%%%%%%%%%%%%%%%%%%%%%%%%%%%%%%%%%%%%%%%%%%%%%%%%%%%%%%%%%%%%%%%%%%

We can now begin to use the VAE architecture to identify the phases of the $J_1$-$J_2$ model as a function of $T$ and $J_2$ for constant $J_1=1$. For $J_2=0$, we are back to the nearest-neighbor Ising model with known critical temperature $T_{c,\text{Ising}} \approx 2.269$ \cite{Onsager}. We can therefore confidently choose at least an initial temperature range of $0 \leq T \leq 4$ containing $T_{c,\text{Ising}}$. From Sec.~\ref{sec:model}, we also know that the ferromagnetic-to-superantiferromagnetic transition is at ${J_2}=1/2$. Hence we choose a range for ${J_2}$ from $0$ to $1.5$ (should we later see that these ranges do not suffice to capture all phases, we 
could
further increase the maximal $T$ and $J_2$ values).
In the present case, we have chosen these $T$ and $J_2$ values to coincide with the range of available data as described in Sec.~\ref{sec:model} where we denoted these ranges as $\mathcal{T}$ and $\mathcal{J}_2$. 

%%%%%%%%%%%%%%%%%%%%%%%%%%%%%%%%%%%%%%%%%%%%%%%%%%%%%%%%%%%%%%%%%%%%%%
\subsection{Global training cycle}
\label{sec:singleVAE-global}

\begin{table}[tb!]
\centering
\begin{tabular}{p{0.3\columnwidth}lD{(}{(}{-1}}%{lD{.}{.}{-1}D{.}{.}{-1}}
\hline
training cycle& loss & \multicolumn{1}{c}{value} \\% & \multicolumn{1}{c}{\textbf{Error}} \\
\hline
%Global Total Loss & 10085.315 & $\pm$ 15.199 \\
global        & $\ell_{\text{KL}}$ & 13.1(3) \\%& \pm 0.3 \\
global        & $\ell_\varepsilon$ & 10072(15) \\[1ex]%& \pm 20 \\
%Ferro Total Loss & 16.665 & $\pm$ 3.175 \\
in-phase  (F) & $\ell_{\text{KL}}$ & 9(2) \\%& \pm 2 \\
in-phase  (F) & $\ell_\varepsilon$ & 8(4) \\%& \pm 4 \\
%Stripe Total Loss & 32.788 & $\pm$ 21.478 \\
in-phase  (S) & $\ell_{\text{KL}}$ & 17(7) \\%& \pm 7 \\
in-phase  (S) & $\ell_\varepsilon$ & 20(20) \\[1ex]%& \pm 20 \\
global        & $\lambda$ & \mathbf{0.2075}(3) \\%& \pm 0.0003 \\
in-phase  (F) & $\lambda$ & \mathbf{0.0012}(2) \\%& \pm 0.0002 \\
in-phase  (S) & $\lambda$ & \mathbf{0.001}(1) \\%& \pm 0.001 \\
\hline
\end{tabular}
\caption{Loss per spin $\lambda$ obtained at epoch $500$ for global and in-phase training cycles as discussed in
Secs.~\ref{sec:singleVAE-global} and \ref{sec:singleVAE-in-phase}, respectively.
The error is calculated as the standard error, given by $\sigma/\sqrt{n}$, where $\sigma$ is the standard deviation of the mean, and $n=100$ is the total number of trainings. The symbols (F) and (S) indicate the ferromagnetic and superantiferromagnetic in-phase training cycles, respectively. The last three (bold) $\lambda$ values have been used in Fig.~\ref{fig:loss-VAE-global}.}
\label{tab:loss_values}
\end{table}

%%%%%%%%%%%%%%%%%%%%%%%%%%%%%%%%%%%%%%%%%%%%%%%%%%%%%%%%%%%%%%%%%%%%%%%%%%%%%%
\begin{figure}[t!]
\begin{center}
\includegraphics[width=1\columnwidth]{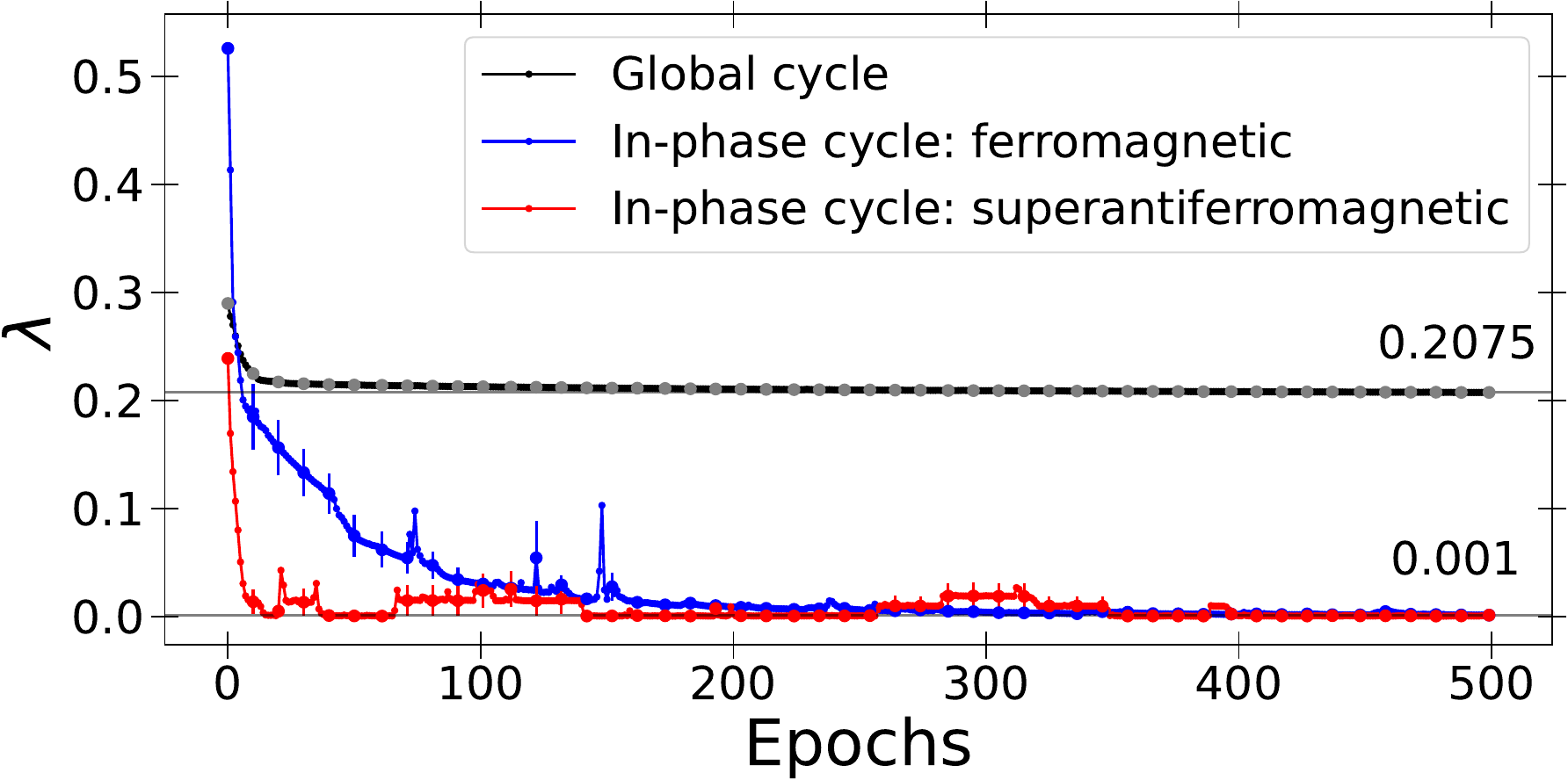}
\end{center}
  \caption{
  Mean loss per site $\lambda$ for $500$ epochs when averaged over $10$ independent trainings. The black markers represent the results for the global training cycle while blue and red represent in-phase training cycles for ferromagnetic training data and superantiferromagnetic training data, respectively. The gray horizontal lines indicate the final mean values of $\lambda$ at epoch $500$. Error bars are the usual standard error of the mean and shown, for clarity, at every $10$th symbol only. The $\lambda$ for the global training cycle is visibly higher than for the in-phase training, due to the selection of $\rho_{G,\text{train}}$ with training samples chosen as specified in Sec.~\ref{sec:singleVAE-global}. 
  }
  \label{fig:loss-VAE-global}
\end{figure}
%%%%%%%%%%%%%%%%%%%%%%%%%%%%%%%%%%%%%%%%%%%%%%%%%%%%%%%%%%%%%%%%%%%%%%

% VAE training
In order to train the \emph{single} VAE, we now choose from each of the $|\mathcal{T}|\times|\mathcal{J}_2|=3454$ pairs $(T,J_2)$ one of the $C$ configurations randomly. In this way, configurations from the whole range of $T$ and $J_2$ values are included in the training cycle. We shall call this dataset $\rho_{G,\text{train}} \subset \rho_{G}$. We train for $500$ epochs with a batch size of $|B|=64$, and achieve a KL loss of $\ell_\text{KL}= 13.1 \pm 0.3$ and an MSE loss of
$\ell_\varepsilon= 10072 \pm 15$, see also the first two lines of Table~\ref{tab:loss_values}.
It may be useful to convert this  to a per-site MSE $\lambda$.
To this end, we first consider the batch size and the number of spins per configuration $30\times30=900$ which results in $3454/64\approx54$ configurations per epoch. The total number of spins in training is thus $54\times900=48600$.
The per-site MSE total loss divided by the total number of spins during the training, is $\lambda={(\ell_\varepsilon+\ell_{\text{KL}})}/{\left( L^2 \frac{|\mathcal{T}|\times|\mathcal{J}_2|}{|B|}\right)}$.
We find that $\ell_\varepsilon$ corresponds to $\lambda=0.2075 \pm 0.0003$ for $L=30$.
As $\lambda$ is a more intuitive metric, we use $\lambda$ to show the evolution of the loss during training epochs in Fig.~\ref{fig:loss-VAE-global}.
One observes good convergence of the training for a total of $500$ epochs.

% VAE testing
The workflow after the training is shown in Fig.~\ref{fig:schematics-flow-diagram}.
We now test how the trained VAE can reconstruct configurations using the test dataset $\rho_{G,\text{test}}$. Here, $\rho_{G,\text{test}}= \rho_G \setminus \rho_{G,\text{train}}$, i.e., the full dataset with $\rho_{G,\text{train}}$ removed.
The size of the test dataset is
$|\rho_{G,\text{test}}|= |\rho_G| - |\rho_{G,\text{train}}| = 138160 - 3454 = 134706$.
For each pair $(T,J_2)$ we then have $C-1=39$ generated spin configurations and can compute $\varepsilon(T,J_2)$ for each. 
In order to use the VAE to reconstruct the phase diagram, we now choose a particular $J'_2 \in \mathcal{J}_2$. We then compute $\varepsilon(T,J'_2)$ for all $T\in \mathcal{T}$.
At each $(T,J'_2)$ there will be a distribution of $39$ $\varepsilon$ values.   
When $T$ and $J_2$ are far away from phase boundaries, the $39$ $\varepsilon(T,J'_2)$ values will follow a roughly similar behavior in each phase. 
On the other hand, close to phase boundaries, there will be a large variation in $\varepsilon(T,J'_2)$. 
One could hence in principle compute $\langle \varepsilon(T,J'_2) \rangle$ to detect a phase change. We have found that $\min_{c \in C}  \varepsilon(T,J'_2)$ works even better for the $J_1$-$J_2$ model.
%
%%%%%%%%%%%%%%%%%%%%%%%%%%%%%%%%%%%%%%%%%%%%%%%%%%%%%%%%%%%%%%%%%%%%%%%%%%%%%%
\begin{figure}[tb!]
\begin{center}
    \includegraphics[width=1\columnwidth]{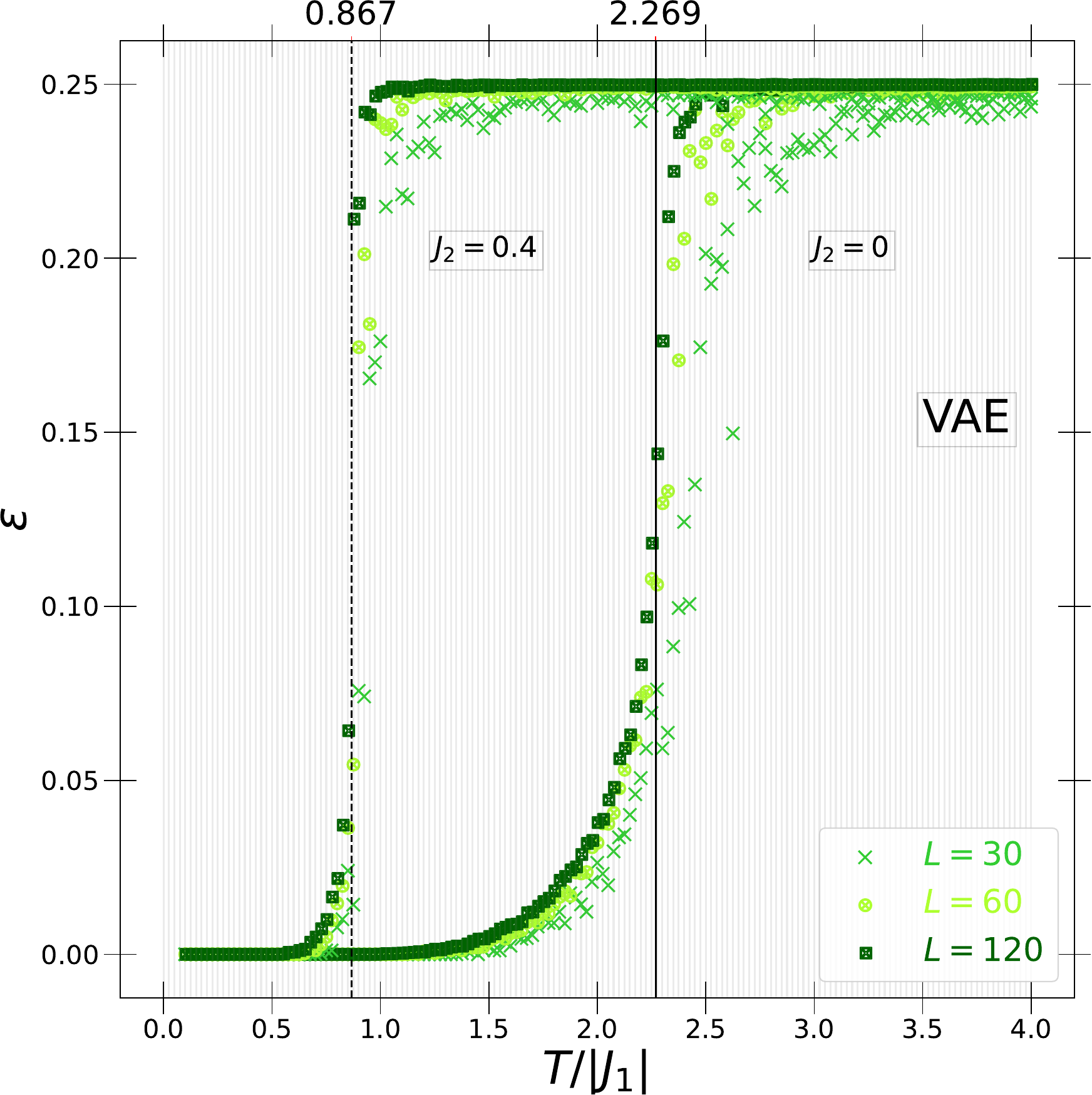}
\end{center}
    \caption{Reconstruction error $\mathcal{E}^{\text{min}}_{\rho_G}(T, \mathcal{J}_2)$ obtained by a single VAE for $J_2=0$ and $0.4$ shown for $L=30$ ($\times$, green), $60$ ($\bigotimes$, light green) and $120$ ($\boxtimes$, dark green). The $157$ gray vertical lines represent the temperatures in $\mathcal{T}$. 
    The black solid vertical line indicates the exact $T_c(J_2=0)=2.269$ and the black dashed line shows the numerical estimate of $T_c(J_2=0.4)=0.867$ from Ref.~\cite{Kalz_2008}.}
    \label{fig:results-VAE-re}
\end{figure}
%%%%%%%%%%%%%%%%%%%%%%%%%%%%%%%%%%%%%%%%%%%%%%%%%%%%%%%%%%%%%%%%%%%%%%

Figure \ref{fig:results-VAE-re} shows results for $J_2= 0$ and $0.4$. 
As expected, for both $J_2$,  we see that $\varepsilon \approx 0$ when $T \ll T_c(J_2)$ while for $T \gg T_c(J_2)$ we find $\varepsilon \approx 0.25$. 
The temperature range where $\varepsilon$ changes is already reasonably close to $T_c(J_2)$ for $L=30$ and we note that, upon increasing $L$,  the curves become sharper with the kink approaching the exact value in the thermodynamic limit.

%%%%%%%%%%%%%%%%%%%%%%%%%%%%%%%%%%%%%%%%%%%%%%%%%%%%%%%%%%%%%%%%%%%%%%
\begin{figure*}[tb]
     \begin{center}
    \raisebox{0\columnwidth}{(a)}\includegraphics[width=0.95\columnwidth]{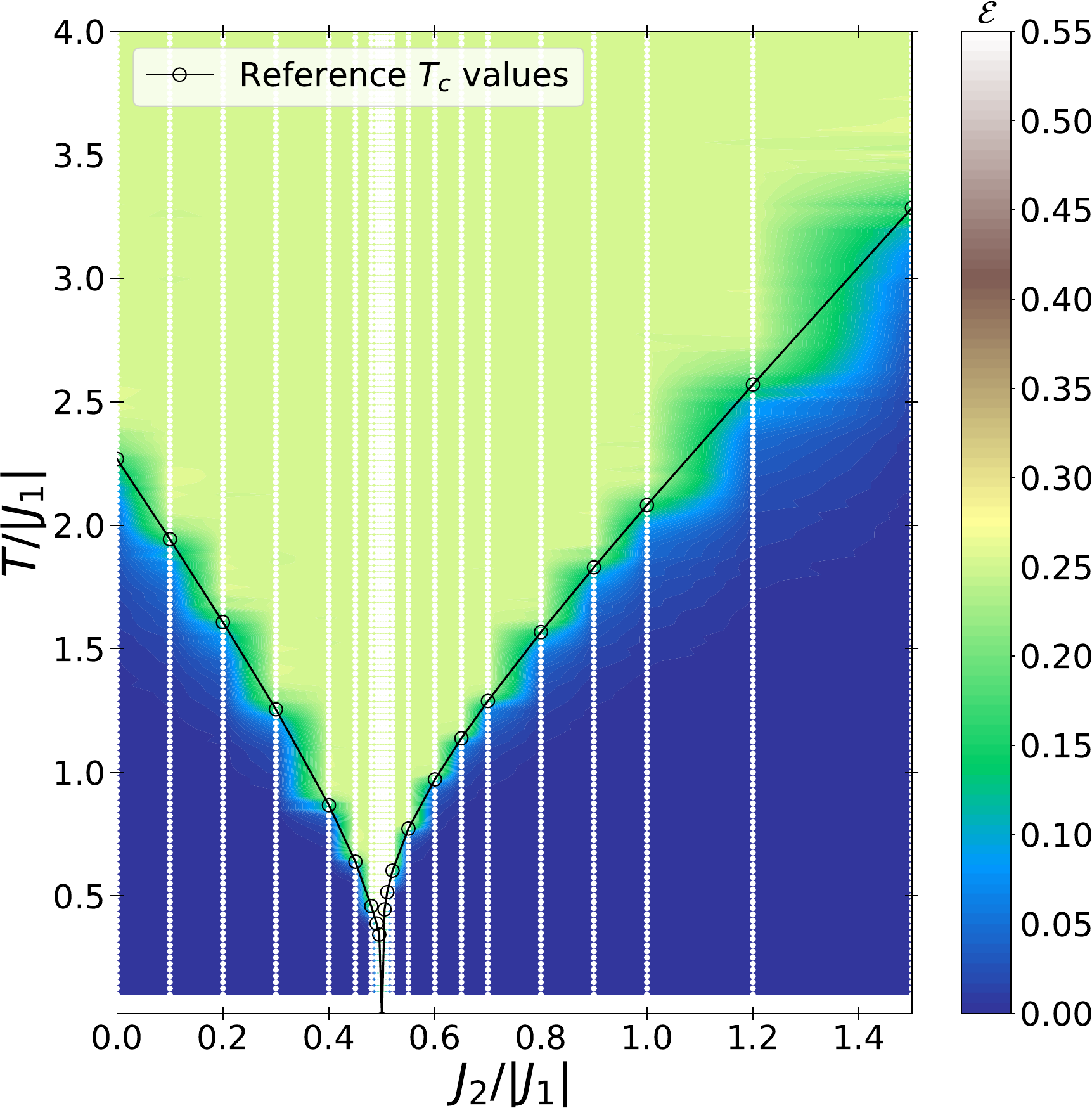}\hfill%
    \raisebox{0\columnwidth}{(b)}\includegraphics[width=0.95\columnwidth]{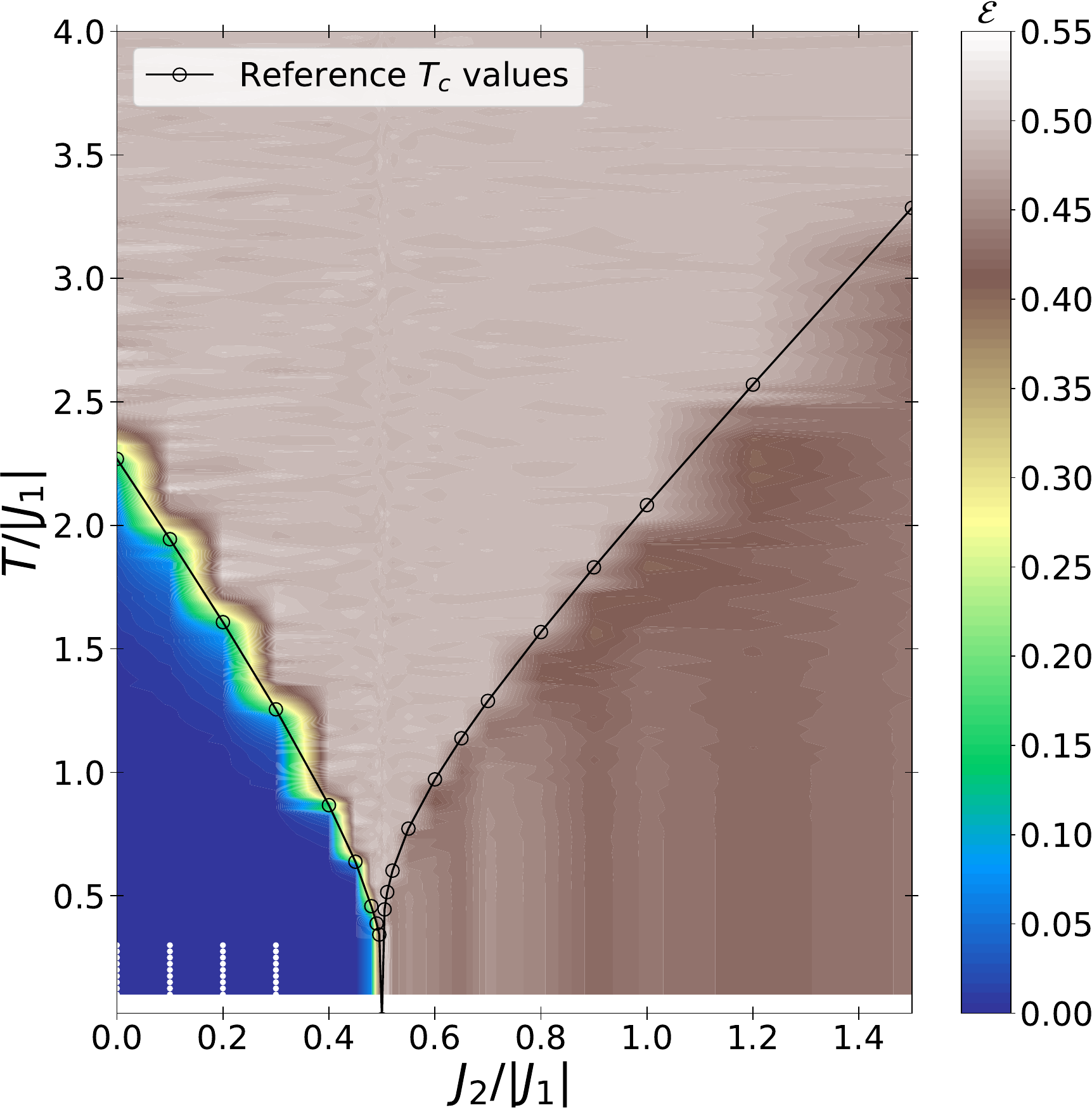}\\ %\hfil%
    \raisebox{0\columnwidth}{(c)}\includegraphics[width=0.95\columnwidth]{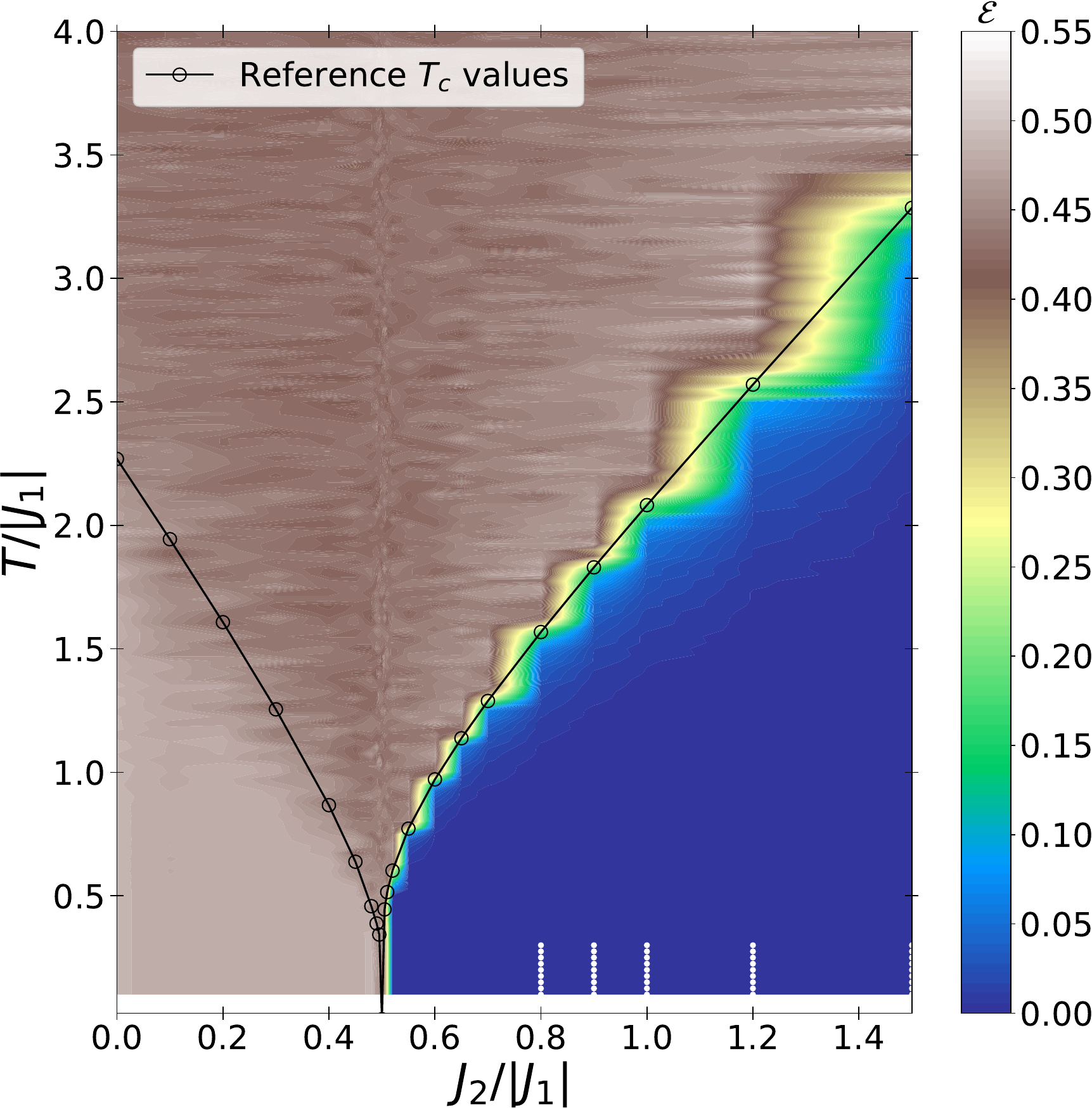}\hfill%
    \raisebox{0\columnwidth}{(d)}\includegraphics[width=0.95\columnwidth]{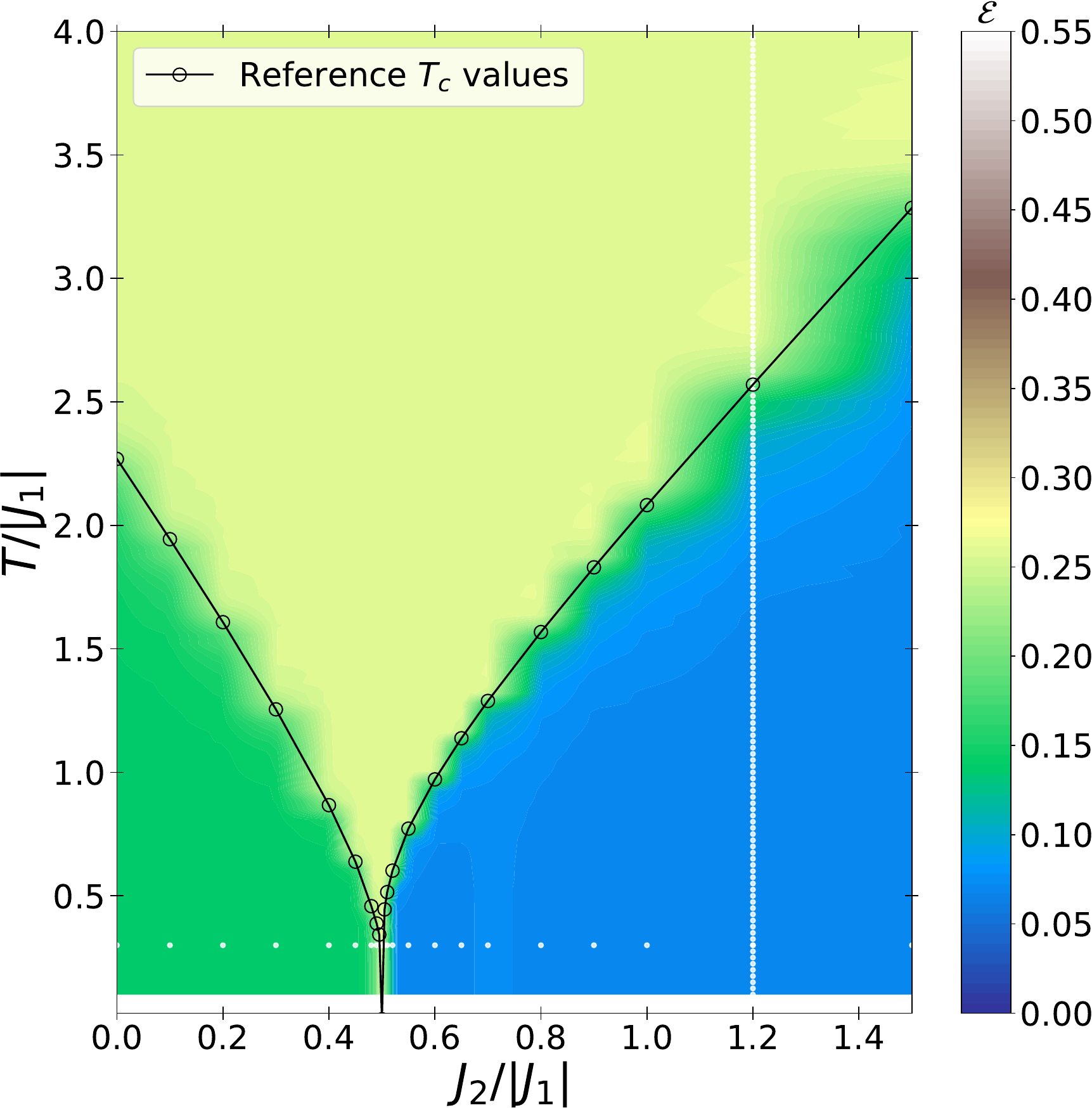}%\hfil%
    \end{center}
%  \vspace*{-1.5mm}
  \caption{$\mathcal{E}^{\text{min}}_{\rho}(\mathcal{T}, \mathcal{J}_2)$ for the VAE-based reconstruction of the $J_1$-$J_2$ model's phase diagram from Secs.~\ref{sec:singleVAE} and \ref{sec:multipleVAE}. The results correspond to $L=30$.
  For the single VAE approach of Sect.~\ref{sec:singleVAE}, panel 
  (a) shows the global training cycle of Sec.~\ref{sec:singleVAE-global} with $\rho=\rho_G$, 
  (b) represents the in-phase learning of Sec.~\ref{sec:singleVAE-in-phase} from the low-$J_2$ region $\rho_{\text{low-}J_2}$ and 
  (c) gives results for the in-phase learning from the high-$J_2$ region $\rho_{\text{high-}J_2}$. The $(T,J_2)$ data points of various training regions are indicated by small white dots for each $(T,J_2)$ pair (usually these are closely spaced and hence appear as vertical lines).
  Lastly, panel 
  (d) shows the results of the multiple VAE approach of Sec.~\ref{sec:multipleVAE}. Here, the white dots give examples for the constant $J_2$ (the vertical line of dots) and the constant $T$ trainings (horizontal line of dots) explained in Sec.~\ref{sec:multipleVAE-constantJ2T}.
  In all panels, $\circ$ symbols connected by black lines denote the reference phase boundaries of Ref.~\cite{Kalz_2008}.}
    \label{fig:results-VAE-phase-diagram}
\end{figure*}
%%%%%%%%%%%%%%%%%%%%%%%%%%%%%%%%%%%%%%%%%%%%%%%%%%%%%%%%%%%%%%%%%%%%%%

Figure~\ref{fig:results-VAE-phase-diagram}(a) shows the emerging phase diagram.
We can see that $\mathcal{E}^{\text{min}}_{\rho_G}(\mathcal{T}, \mathcal{J}_2)$ separates into two distinct regions. Let us emphasize that up to this point, we have not used any a priori information besides the two limiting transition temperatures as outlined below. In particular, we have not yet used information about the spatial distribution of the spins in each of the configurations. It is hence noteworthy to find that the border between the two identified regions already is very close to the known phase boundaries.
The apparent steps in the VAE estimates for the phase boundaries arise in fact from the underlying discrete set of values of $T$ and in particular $J_2$ investigated whereas the agreement for actual values, as denoted by the circles, is in fact at the resolution limit of the figure.

%%%%%%%%%%%%%%%%%%%%%%%%%%%%%%%%%%%%%%%%%%%%%%%%%%%%%%%%%%%%%%%%%%%%%%
\subsection{In-phase training cycles}
\label{sec:singleVAE-in-phase}

Having identified two distinct regions in Sec.~\ref{sec:singleVAE-global}, we can now repeat the training of the VAE deep in one of the two phases. In Figs.~\ref{fig:results-VAE-phase-diagram}(b,c) we indicate two distinct such regions for low $T$, following the general shape of the boundary between the two regions which seems to have a clear separation into low- and high-$J_2$ sub-regions.
We follow the same procedure as for Fig.\ \ref{fig:results-VAE-phase-diagram}(a), but with the much more restricted training data regions
$\rho_{\text{low-}J_2}$ and $\rho_{\text{high-}J_2}$. In order to have a reasonable amount of training data, we now use all $40$ values for each $(T, J_2)$ in each training region. For the results underlying  Fig.~\ref{fig:results-VAE-phase-diagram}(b), this amounts to $1440$ training configurations in $\rho_{\text{low-}J_2}$, while for  Fig.~\ref{fig:results-VAE-phase-diagram}(c), we have $1800$ configurations in $\rho_{\text{high-}J_2}$. Loss function convergence values are presented in Table~\ref{tab:loss_values}.

From  Fig.~\ref{fig:results-VAE-phase-diagram}(b) we see that again two distinct regions emerge. Now, the low-$T$, low-$J_2$ region is clearly separated from the rest of the $(T, J_2)$ plane. Similarly, Fig.~\ref{fig:results-VAE-phase-diagram}(c) establishes a low-$T$, high-$J_2$ region.
We note that in both cases, the $\varepsilon$ values in the low/high-$J_2$ regions are close to zero, while in the other regions we have $\varepsilon\approx 0.5$.
This value suggests that in both cases, the out-of-region configurations have about $50\%$ of spins different, in agreement with the known phases as presented in Fig.~\ref{fig:phase-diagram}.
We can therefore conclude that the low-$T$ region identified in Fig.~\ref{fig:results-VAE-phase-diagram}(a) consists of two distinct regions.

%%%%%%%%%%%%%%%%%%%%%%%%%%%%%%%%%%%%%%%%%%%%%%%%%%%%%%%%%%%%%%%%%%%%%%
\subsection{Discussion of reconstruction error}
\label{sec:singleVAE-re}

Overall, the combination of global and in-phase learning has indeed led to the identification of three separate regions. These regions agree very well with the previously established phases shown in Fig.~\ref{fig:phase-diagram}. The $\varepsilon$ values of $0$, $0.25$, and $0.5$ indicate best, random, and worst reconstruction possible, respectively, compatible with the spin configurations in each phase. 
Clearly, the regions with $\varepsilon \approx 0$ correspond to the ordered ferro- and superantiferromagnetic phases.
When using the VAEs trained in both of these phases, we find $\varepsilon \approx 0.5$ when testing in the disordered paramagnetic phase, clearly showing the difference between ordered and disordered phases.

The value of $\varepsilon \approx 0.25$, however, should not emerge when simply comparing the configuration of up ($+1$) and down ($-1$) spin states as shown in Figs.~\ref{fig:sample-configurations}(a--e). However, it is compatible with the VAE reconstruction of spin configurations similar to the undifferentiated ``gray'' in Figs.~\ref{fig:sample-configurations}(g+h). 
This tells us that even though we can train the VAEs in the disordered phase, this does not lead to any predictive power even in the disordered phase. Put differently: if we had chosen a third region to train our VAEs, namely a high-$T$ region in the disordered phase, we would not have been able to identify the phase boundaries to the two ordered phases.

%%%%%%%%%%%%%%%%%%%%%%%%%%%%%%%%%%%%%%%%%%%%%%%%%%%%%%%%%%%%%%%%%%%%%%
\section{Reconstruction of the phase diagram using multiple VAEs}
\label{sec:multipleVAE}
%%%%%%%%%%%%%%%%%%%%%%%%%%%%%%%%%%%%%%%%%%%%%%%%%%%%%%%%%%%%%%%%%%%%%%

Another approach is training \emph{multiple} VAEs for different regions $\rho$ spanning across the $(T,J_2)$ plane. 
Let $\rho_{\mathcal{T}'}(J_2)$ denote a region at constant $J_2$ with varying temperature in
$\mathcal{T}'= \{ 0.1, \ldots, 4 \}$. 
$\mathcal{T}'$ consists of $40$ evenly spaced temperatures with $\Delta T= 0.1$ and $\mathcal{T}' \subset \mathcal{T}$. 
In Fig.~\ref{fig:results-VAE-phase-diagram}(d), we indicate as example $\rho_{\mathcal{T}'}(J_2=1.2)$. 
Additionally, we define by $\rho_{\mathcal{J}_2}(T)$ regions of constant $T$ but containing the $22$ elements of $\mathcal{J_2}$. An example of such a region is again given in Fig.~\ref{fig:results-VAE-phase-diagram}(d), namely for $T=0.25$.

Then for a given $(T,J_2)$ pair, we compute $\varepsilon_{\rho}(T,J_2)$. 
In order to remove the bias of low $\varepsilon$ for in-region trainings, we average over the reconstruction errors computed from all $\rho$, i.e., 
\begin{equation}
    \langle \varepsilon \rangle(T,J_2)= 
    \frac{
    \sum_{\rho \in \rho_{\mathcal{T}'}(J_2), \rho_{\mathcal{J}_2}(T) } \varepsilon_{\rho}(T,J_2)
    }{
    |\rho_{\mathcal{T}'}(J_2)| + |\rho_{\mathcal{J}_2}(T)|
    } .
\end{equation}
As we shall show below, this also allows the identification of three separate regions. Although the method corresponds to training $|\mathcal{J}_2| + |\mathcal{T}'|= 62$ VAEs as outlined below, this is in practice not much more involved than the single-VAE method of Sec.~\ref{sec:singleVAE} and can, arguably, be seen as less biased since we do not select a-priori specific regions to train for (cf.\ Sec.~\ref{sec:singleVAE-in-phase}). 
Before proceeding, let us streamline the notation again and denote the training set of the $62$ VAEs by $\rho_{\Delta G}$.

%%%%%%%%%%%%%%%%%%%%%%%%%%%%%%%%%%%%%%%%%%%%%%%%%%%%%%%%%%%%%%%%%%%%%%
\subsection{Constant $J_2$ and $T$ training cycles}
\label{sec:multipleVAE-constantJ2T}

%%%%%%%%%%%%%%%%%%%%%%%%%%%%%%%%%%%%%%%%%%%%%%%%%%%%%%%%%%%%%%%%%%%%%%%%%%%%%%
\begin{figure}[tb!]
\begin{center}
    \includegraphics[width=1\columnwidth]{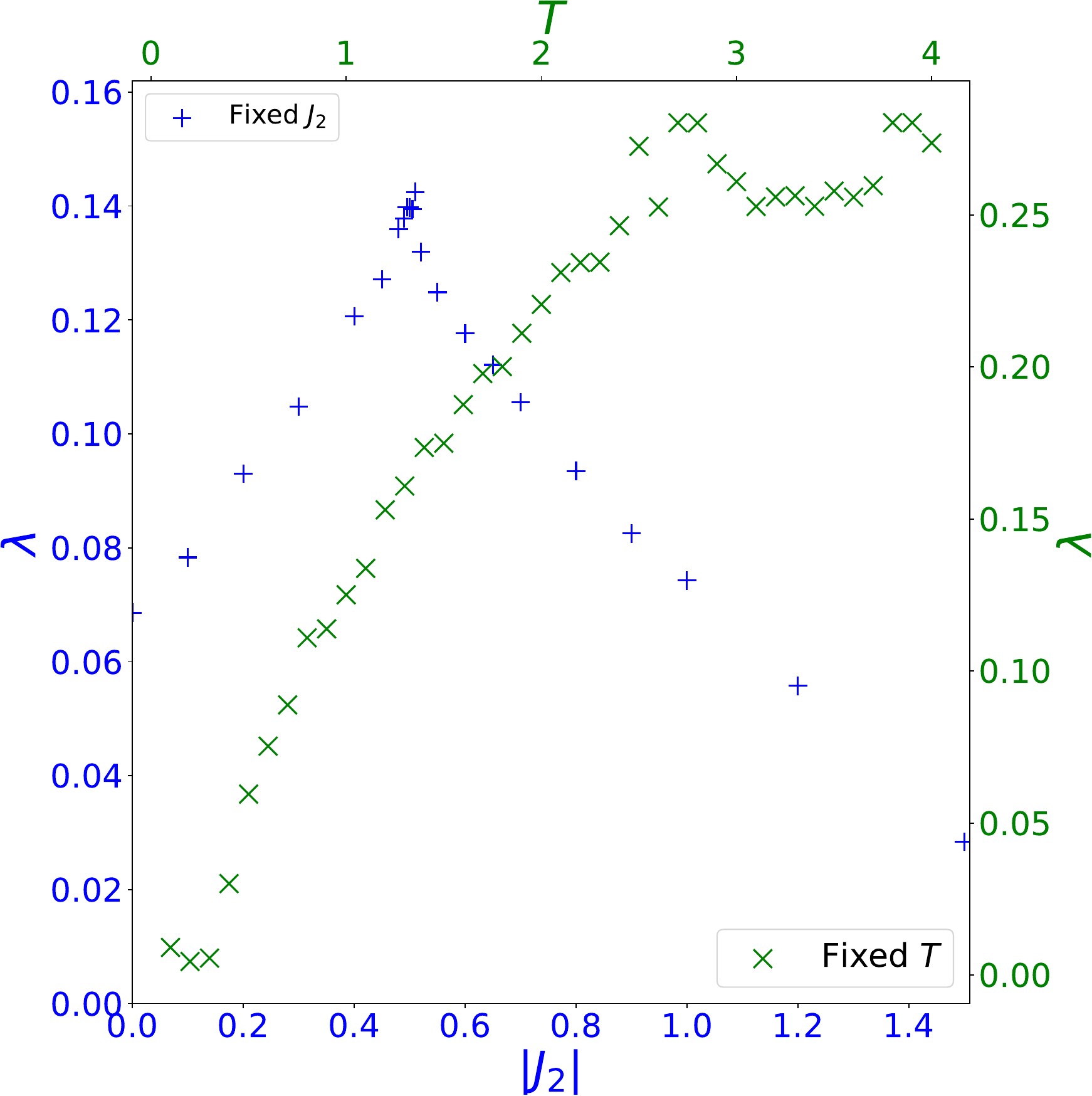}
\end{center}
    \caption{Loss per spin $\lambda$ obtained at epoch $500$ for multi-VAE training cycles as discussed in Sec.~\ref{sec:multipleVAE}. Blue $(+)$ represents the loss per spin $\lambda$ at fixed $J_2$ training cycle, and green $(\times)$ are the losses $\lambda$ for a fixed $T$ training cycle.
    }
    \label{fig:multiVAE-j2}
\end{figure}
%%%%%%%%%%%%%%%%%%%%%%%%%%%%%%%%%%%%%%%%%%%%%%%%%%%%%%%%%%%%%%%%%%%%%%

In the first instance, we train $22$ VAEs, corresponding to each of the $22$ $J_2\in \mathcal{J}_2$. The training is done for each VAE on the $40$ temperatures in $\mathcal{T}'$ outlined above, using the $C=40$ available configurations at each $(T,J_2)$. In Fig.\ \ref{fig:results-VAE-phase-diagram}(d), we indicate this by the vertical line of white dots at $J_2=1.2$; the other $21$ lines for the remaining $J_2$ values are not shown for clarity.
Next, we train additional VAEs by selecting a fixed $T$ among the possible values $0, 0.1, \ldots, 4$ and train a VAE with varying $J_2$ values. In Fig.\ \ref{fig:results-VAE-phase-diagram}(d), this is indicated this by the horizontal line of white dots at $T=0.25$. Since we have $40$ available $T$'s to use, this results in $40$ further trained VAEs.

Figure \ref{fig:multiVAE-j2} shows
the corresponding losses $\lambda$ at the end of a training cycle.
The precise value of $\lambda$ depends on details of the training cycle such as the number of configurations used. 
Nevertheless, at a qualitative level, we can interpret these results as follows. Deep inside an ordered phase, the VAE learns configurations well such that the loss $\lambda$ is close to $0$. On the other hand, the VAE is unable to learn a disordered configuration and rather returns an average gray for these, compare Fig.~\ref{fig:sample-configurations}(g+h).
This amounts to a loss $\lambda \approx 0.25$. When the training dataset contains a mixture of ordered and disordered configurations,
the overall loss seems to be a weighted average of these two limits. The results in Fig.~\ref{fig:multiVAE-j2} thus reflect the fraction of the disordered configurations in the corresponding cut at fixed $J_2$ or $T$, respectively.

%%%%%%%%%%%%%%%%%%%%%%%%%%%%%%%%%%%%%%%%%%%%%%%%%%%%%%%%%%%%%%%%%%%%%%
\subsection{Result for averaged VAEs}
\label{sec:multipleVAE-averages}

Armed with the $62$ trained VAEs, we can now go to each of the $|\mathcal{T}'| \times |\mathcal{J}_2| = 157 \times 22$ points in the $(T,J_2)$ plane, compute $\varepsilon_{\rho_{\mathcal{T}'}}(T,J_2)$ for each of the $40$ configurations and then average to create $\mathcal{E}_{\rho_{\mathcal{T}'}}(\mathcal{T},\mathcal{J}_2)$. 
The result is shown in Fig.~\ref{fig:results-VAE-phase-diagram}(d). As in the other panels of Fig.~\ref{fig:results-VAE-phase-diagram}, we find a distinction between
different
phases. The changes from one phase to the next, when plotted for constant $J_2$, are very similar to Fig.~\ref{fig:results-VAE-re}.
We find 
that there are three distinct regions, namely one with $\varepsilon \approx 0$, a second one with $\approx 0.25$ and the third showing $\varepsilon \approx 0.17$. These regions match the superantiferromagnetic, the ferromagnetic, and the paramagnetic phases in Fig.\ \ref{fig:phase-diagram} very well.

%%%%%%%%%%%%%%%%%%%%%%%%%%%%%%%%%%%%%%%%%%%%%%%%%%%%%%%%%%%%%%%%%%%%%%
\section{Constructing the phase diagram by comparing configurations}
\label{sec:ic}
%%%%%%%%%%%%%%%%%%%%%%%%%%%%%%%%%%%%%%%%%%%%%%%%%%%%%%%%%%%%%%%%%%%%%%

Now we will investigate if one can shortcut the technical complications of the VAE and perform a direct comparison of configurations (CMP) instead.
We proceed analogously to Sec.~\ref{sec:singleVAE-global} and again select one of the $C$ configurations randomly from each of the $|\mathcal{T}|\times|\mathcal{J}_2|=3454$ pairs $(T,J_2)$ shown in Fig.\ \ref{fig:results-VAE-phase-diagram}. We shall denote this reference set as $\varrho_G$. Instead of training as in the VAE, the CMP approach
simply computes an averaged reconstruction error $\langle\varepsilon_{\varrho_G}\rangle(T,J_2)$ for a given test pair $(T,J_2)$, where the mean is found by averaging over all $\varepsilon$ for each data point in $\varrho_G$.
In order to simplify again the notation, and since the average is somewhat implicit in the use of the subscript $\varrho_G$, we shall proceed by
using $\varepsilon_{\varrho_G}(T,J_2)$ to denote this averaged reconstruction error.

%%%%%%%%%%%%%%%%%%%%%%%%%%%%%%%%%%%%%%%%%%%%%%%%%%%%%%%%%%%%%%%%%%%%%%
\subsection{Global comparison cycle}
\label{sec:ic-global}

\begin{figure}
\centering
   \includegraphics[width=1\columnwidth]{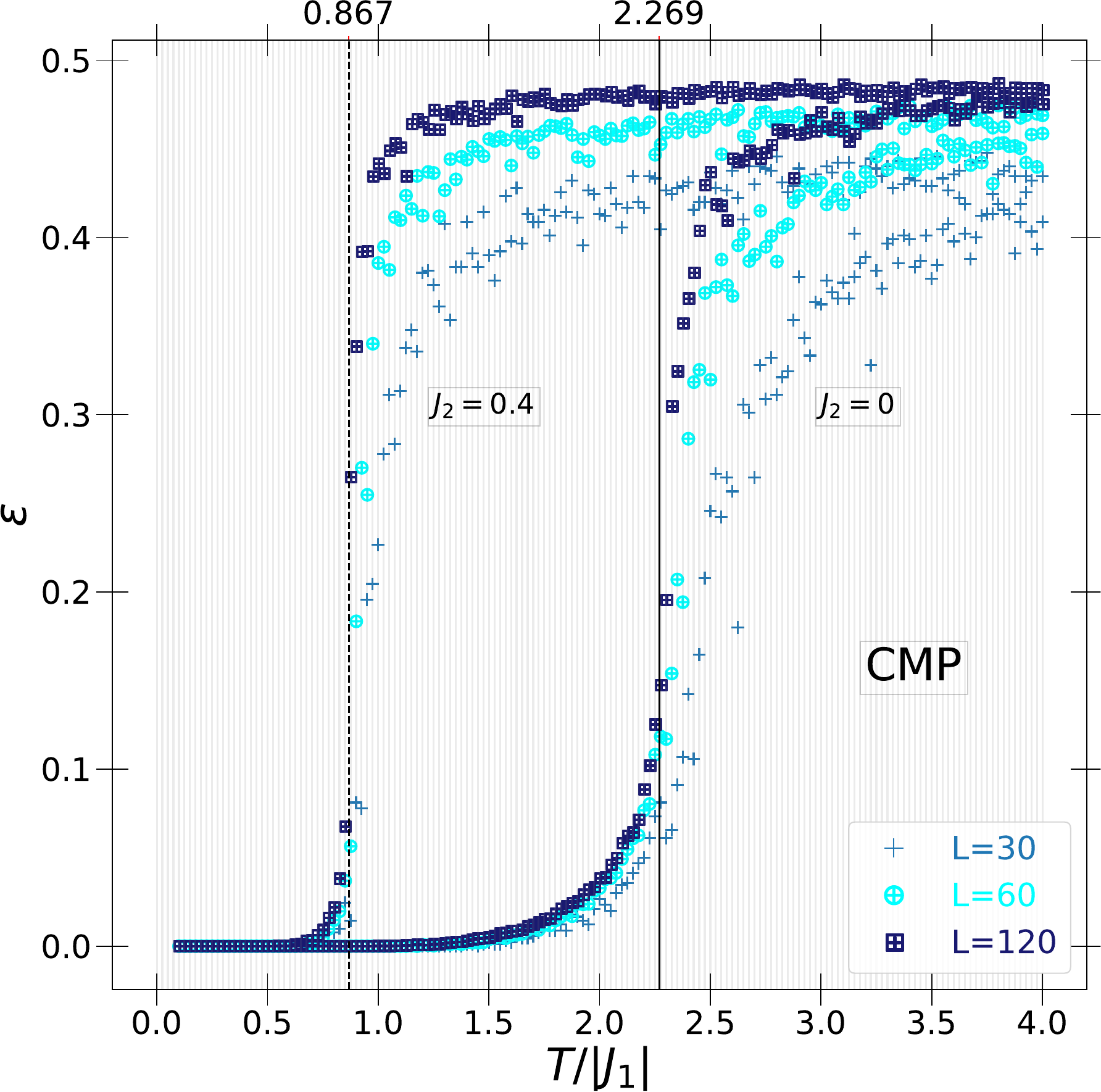}
   \caption{
   Reconstruction error $\mathcal{E}^{\text{min}}_{\rho_G}(T, \mathcal{J}_2)$ obtained by direct comparison of configurations for $J_2=0$ and $0.4$ shown for $L=30$ ($+$, blue), $L=60$ ($\bigoplus$, light blue), and $L=120$ ($\boxplus$, dark blue).
   As in Fig.\ \ref{fig:results-VAE-re}, the vertical gray and black (solid and dashed) lines show the temperature resolution and positions of $T_c$ for $J_2=0$ and $0.4$, respectively.}
   \label{fig:results-IC-re}
\end{figure}

As in Sec.~\ref{sec:singleVAE-global}, we have $39$ data points available to compute $\varepsilon_{\varrho_G}(T,J_2)$ for each pair $(T,J_2)$. We again choose a $J'_2 \in \mathcal{J}_2$ and compute $\varepsilon_{\varrho_G}(T,J'_2)$ for all $T \in \mathcal{T}$. This yields $39$ values for each $(T,J_2)$. Studying again $\varepsilon^{\text{min}}_{\varrho_G}(T,J'_2)$, Figure \ref{fig:results-IC-re} 
shows results for ${J'_2}=0$ and $0.4$.
The result is qualitatively remarkably close to that of Fig.~\ref{fig:results-VAE-re}, showing a change from low $\varepsilon^{\text{min}}_{\varrho_G} \sim 0$ to values close to $0.5$ at high temperatures. The values of $\varepsilon^{\text{min}}_{\varrho_G}$ are very similar and as before, going from $L=30$ to $60$,  and finally to $120$ makes the change of $\varepsilon_{\varrho_G}$ more pronounced when passing from the low-$T$ region to the high-$T$ one.
Figure \ref{fig:results-IC-phase-diagram}(a) presents an overview of the results of this approach for the full $J_2$-$T$ regime for the $L=30$ system. Overall, we can distinguish two regions: a low-temperature one with low reconstruction error $\varepsilon$ and a high-temperature one where $\varepsilon$ approaches $0.5$.

%%%%%%%%%%%%%%%%%%%%%%%%%%%%%%%%%%%%%%%%%%%%%%%%%%%%%%%%%%%%%%%%%%%%%%
\subsection{In-phase comparison cycle}
\label{sec:ic-in-phase}

\begin{figure*}[tb!]
     \begin{center}
    \raisebox{0\columnwidth}{(a) }\includegraphics[width=0.95\columnwidth]{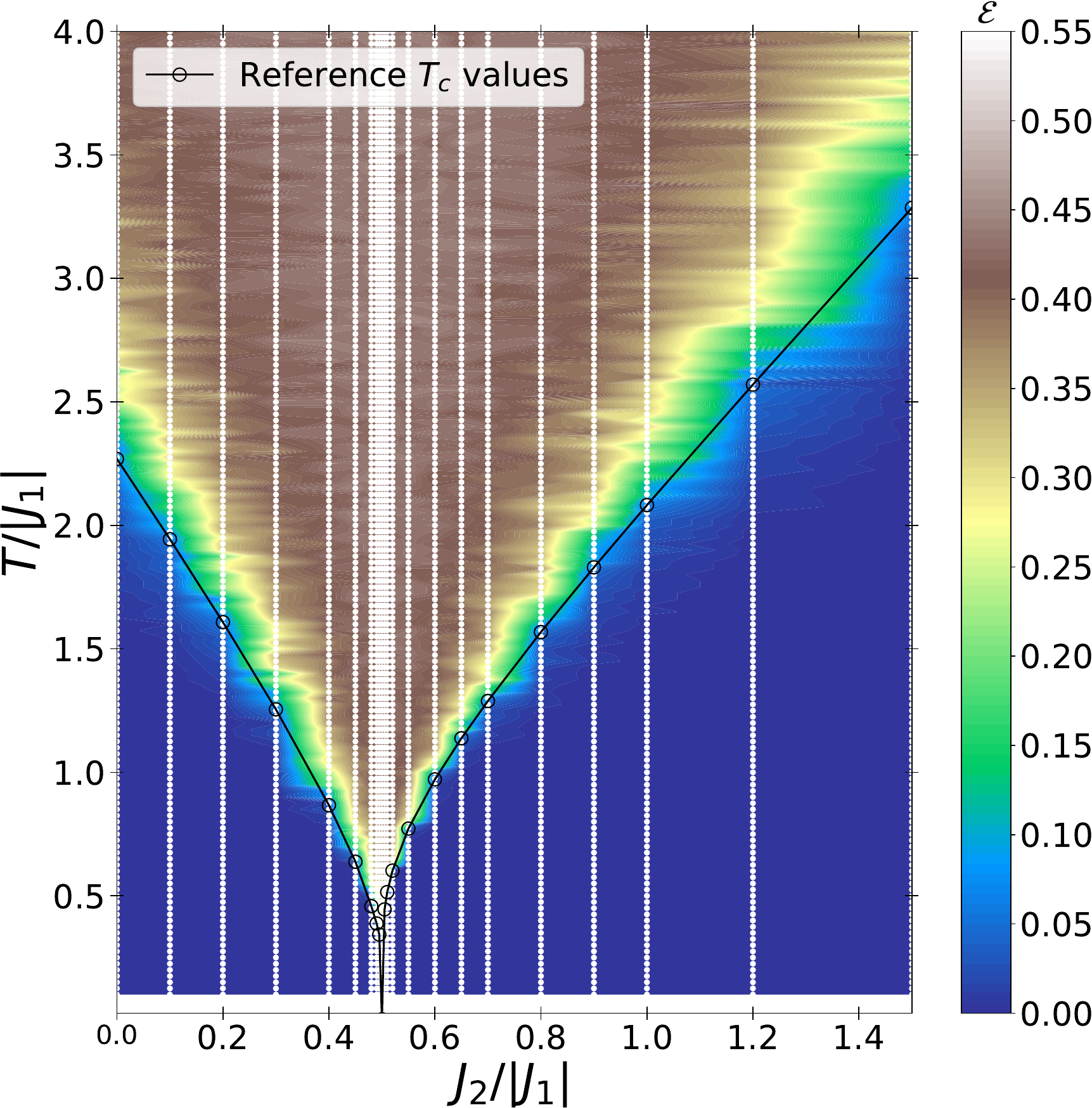}\hfill%
    \raisebox{0\columnwidth}{(b) }\includegraphics[width=0.95\columnwidth]{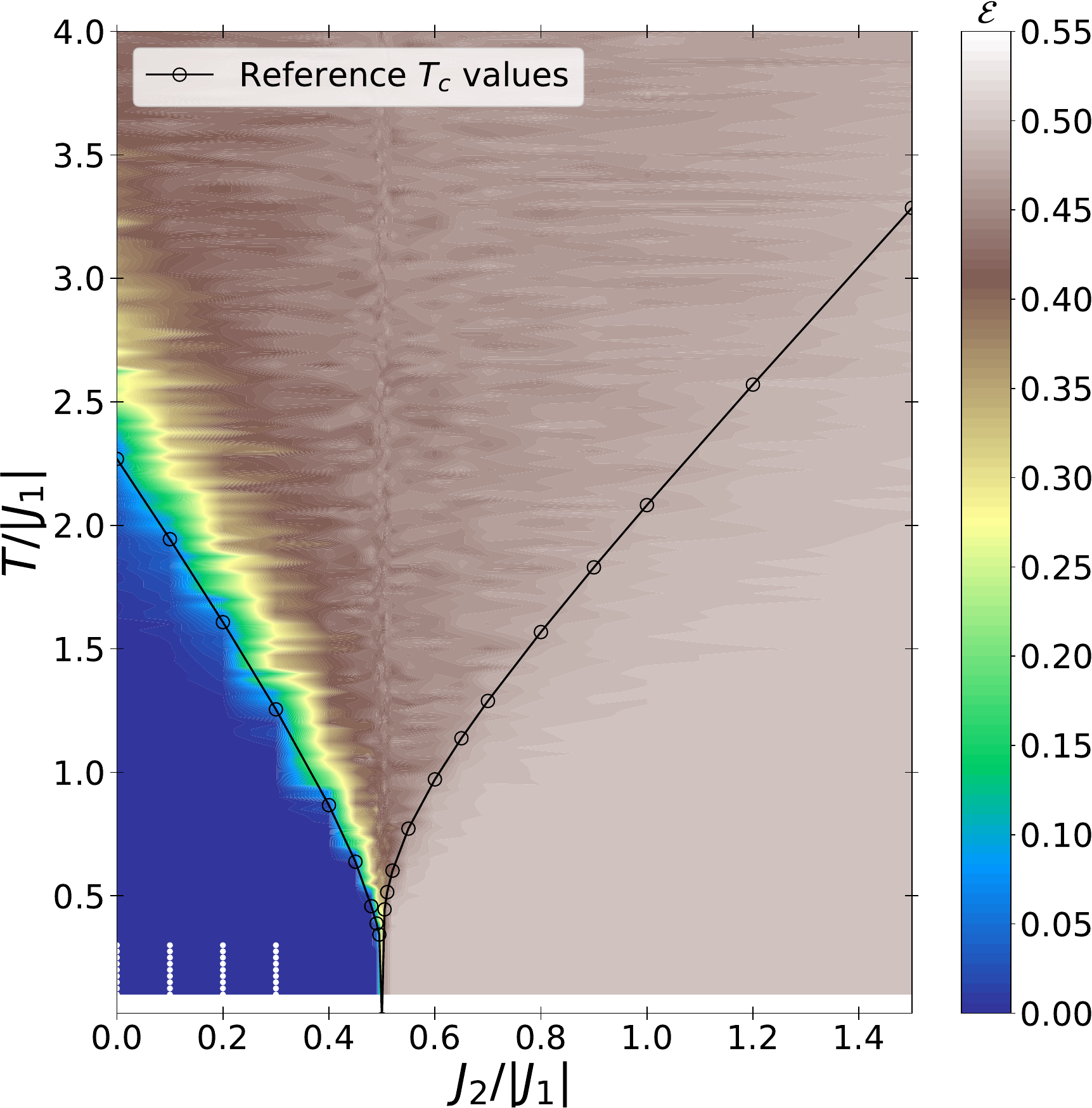}\\%\hfil%
    \raisebox{0\columnwidth}{(c) }\includegraphics[width=0.95\columnwidth]{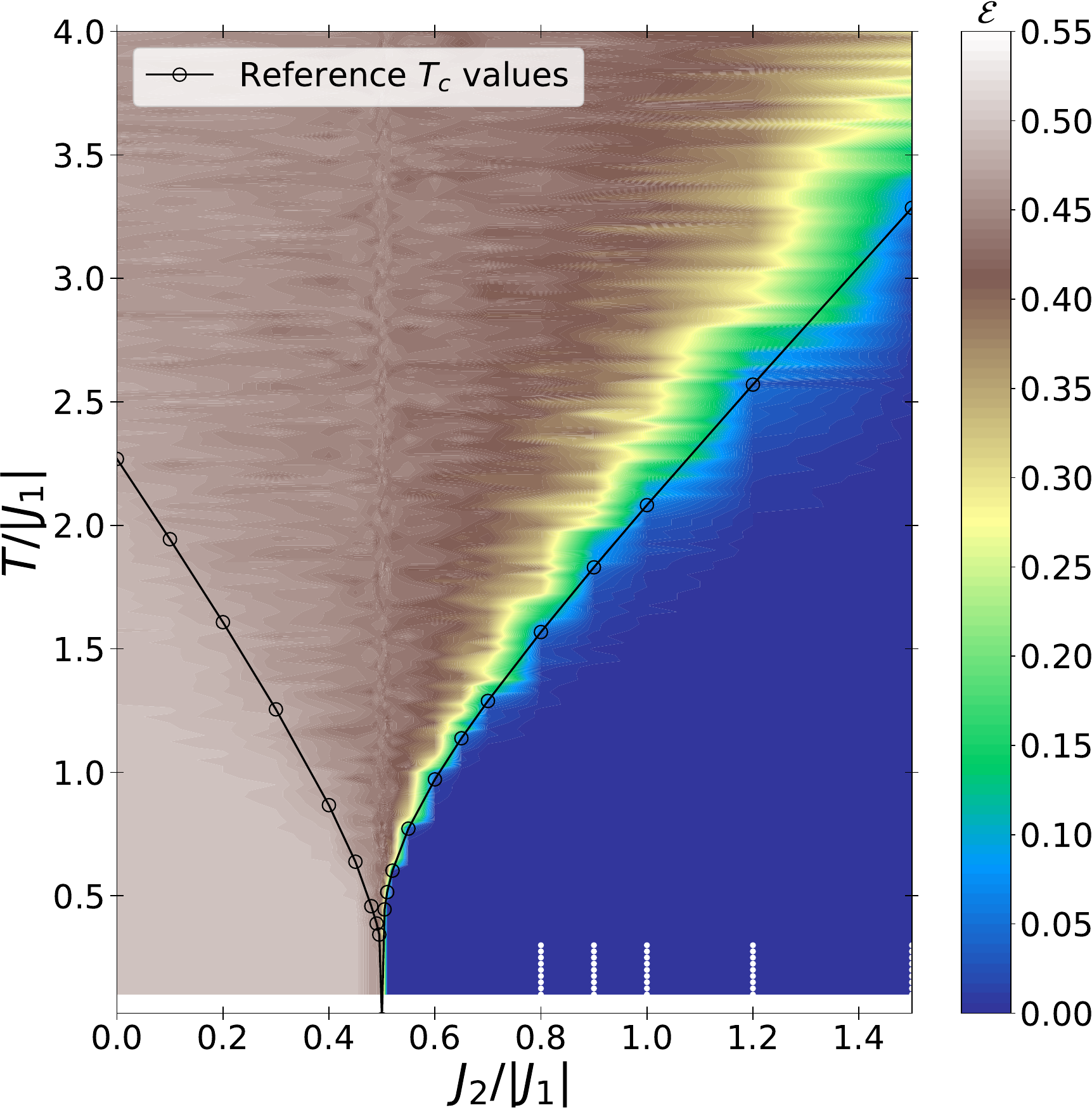}\hfill%
    \raisebox{0\columnwidth}{(d) }\includegraphics[width=0.95\columnwidth]{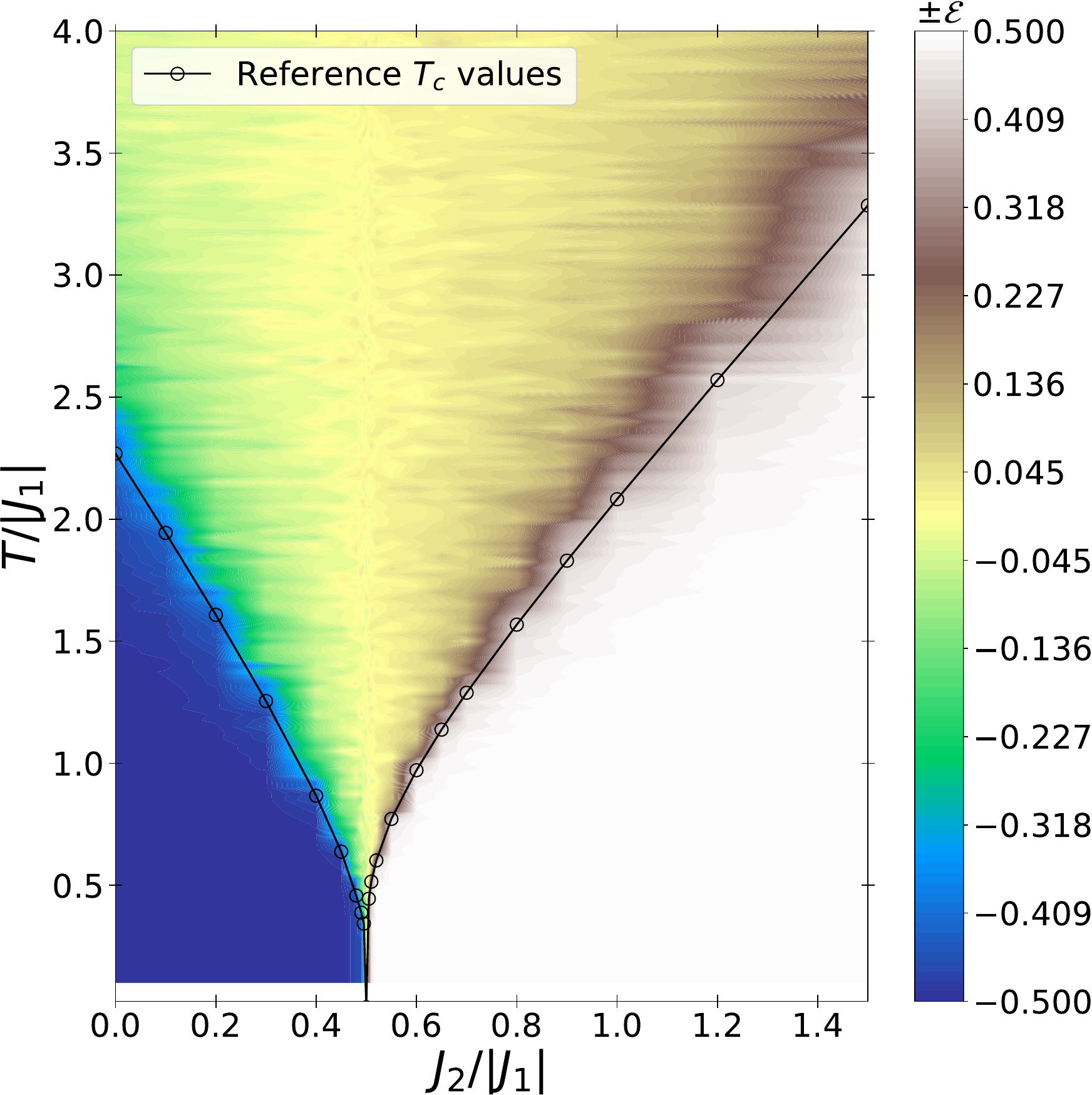}%\hfil%
    \end{center}
  \caption{
  Reconstruction error $\mathcal{E}^{\text{min}}_{\rho_G}(\mathcal{T}, \mathcal{J}_2)$ for the CMP-based reconstruction of the phase diagram for the $J_1$-$J_2$ Ising model with $L=30$.
  (a) Contour plot of all the $\mathcal{E}_{\text{CMP}}(J_2)$ computed with the global comparison cycle, described in Sec.~\ref{sec:ic-global}, made by randomly selecting reference configurations
  from the entire dataset. 
  (b) Contour plot of all the $\mathcal{E}_{\text{CMP}}(J_2)$ computed with the in-phase comparison cycle, described in Sec.~\ref{sec:ic-in-phase}, made by choosing reference configurations from $(T,J_2)$ tuples such that $J_2\in\{0.0,0.1,0.2,0.3\}$ and $T\in\{0.1\dots0.3\}$, also represented as white points. 
  (c) Contour plot of all the $\mathcal{E}_{\text{CMP}}(T,J_2)$ computed with the in-phase comparison cycle, made by choosing reference configurations from $(T,J_2)$ tuples such that $J_2\in\{0.8,0.9,1.0,1.1,1.2,1.3,1.4,1.5\}$ and $T\in\{0.1\dots0.3\}$ (\ref{sec:ic-in-phase}). 
    (d) Element-wise difference of (b) and (c), i.e., (b)$-$(c).
    As in Fig.~\ref{fig:results-VAE-phase-diagram}, the $\circ$ symbols connected by black lines denote the reference phase boundaries of Ref.~\cite{Kalz_2008} in each panel.}
    \label{fig:results-IC-phase-diagram}
\end{figure*}

Following the reasoning of Sec.~\ref{sec:ic-in-phase}, we again define two low-$T$ regions as indicated in Figs.~\ref{fig:results-IC-phase-diagram}(b+c). We equip the CMP with the restricted data region, e.g., $\varrho_{\text{low-}J_2}$ and $\varrho_{\text{high-}J_2}$, and as before use the full $40$ spin configurations for each $(T,J_2)$ in $\varrho_{\text{low-}J_2}$ and $\varrho_{\text{high-}J_2}$. 
Figure \ref{fig:results-IC-phase-diagram}(b) shows that two regions can be identified; a low-$T$, low-$J_2$ region has separated from the rest of the $(T,J_2)$ plane. For Fig.\ \ref{fig:results-IC-phase-diagram}(c), a similar observation allows to differentiate a low-$T$, high-$J_2$ region in the $(T,J_2)$ plane.
The $\varepsilon$ values in the low/high-$J_2$ regions in  Figs.~\ref{fig:results-IC-phase-diagram}(b+c) are close to zero, while in the other regions we have $\varepsilon\approx 0.5$.
The results for the CMP approach are hence also in good agreement with the known phases as presented in Fig.\ \ref{fig:phase-diagram} and we can therefore again conclude that the low-$T$ region identified in Fig.\ \ref{fig:results-IC-phase-diagram}(a) consists of two distinct regions.

We can combine Figs.~\ref{fig:results-IC-phase-diagram}(b+c) into one by considering their element-wise difference.
Figure \ref{fig:results-IC-phase-diagram}(d) shows the resulting phase diagram where, as in
Fig.~\ref{fig:results-VAE-phase-diagram}(d),
all three phases can be distinguished clearly.

%%%%%%%%%%%%%%%%%%%%%%%%%%%%%%%%%%%%%%%%%%%%%%%%%%%%%%%%%%%%%%%%%%%%%%
%%%%%%%%%%%%%%%%%%%%%%%%%%%%%%%%%%%%%%%%%%%%%%%%%%%%%%%%%%%%%%%%%%%%%%

%%%%%%%%%%%%%%%%%%%%%%%%%%%%%%%%%%%%%%%%%%%%%%%%%%%%%%%%%%%%%%%%%%%%%%
%%%%%%%%%%%%%%%%%%%%%%%%%%%%%%%%%%%%%%%%%%%%%%%%%%%%%%%%%%%%%%%%%%%%%%

%%%%%%%%%%%%%%%%%%%%%%%%%%%%%%%%%%%%%%%%%%%%%%%%%%%%%%%%%%%%%%%%%%%%%%
\subsection{Discussion of reconstruction error}
\label{sec:ic-re}

The VAE returned a reconstruction error $\varepsilon \approx 0.25$ in the disordered high-temperature phase (compare Figs.~\ref{fig:results-VAE-re}
and \ref{fig:results-VAE-phase-diagram}(a+d)).
The CMP procedure yields instead a twice larger value $\varepsilon \approx 0.5$ (see Figs.~\ref{fig:results-IC-re} and
\ref{fig:results-IC-phase-diagram}).
The explanation is very simple: while the VAE reproduces the gray images shown in Figs.~\ref{fig:sample-configurations}(g+h) for the disordered phase, the direct calculation of $\varepsilon$ only uses the
discrete $\pm 1$ spin values.
Thus, while the $\varepsilon=0.25$ of the VAE corresponds to the squared distance of either black or white pixels to an average gray,
the value $\varepsilon=0.5$ found in the CMP procedure corresponds to the average of half of the pixels in two random images being identical and the other half is the opposite of each other. 
We also note that in Fig.\ \ref{fig:results-IC-phase-diagram}, one has to be quite far from the phase boundaries to find $\varepsilon\approx 0.5$ to a good accuracy while mostly $\varepsilon \lesssim 0.5$. This behavior suggests that the
configuration comparison is sensitive to the difference of each spin in any two configurations being compared. Close to the phase boundary, the system undergoes a rapid change between two phases, causing a rapid change in the values of spin representation, resulting in $\varepsilon \lesssim 0.5$.  
%%%%%%%%%%%%%%%%%%%%%%%%%%%%%%%%%%%%%%%%%%%%%%%%%%%%%%%%%%%%%%%%%%%%%%
\section{Identifying the phase boundaries}
%%%%%%%%%%%%%%%%%%%%%%%%%%%%%%%%%%%%%%%%%%%%%%%%%%%%%%%%%%%%%%%%%%%%%%

%%%%%%%%%%%%%%%%%%%%%%%%%%%%%%%%%%%%%%%%%%%%%%%%%%%%%%%%%%%%%%%%%%%%%%
\begin{figure*}[tb]
     \begin{center}
    \raisebox{0\columnwidth}{(a)}\includegraphics[width=0.95\columnwidth]{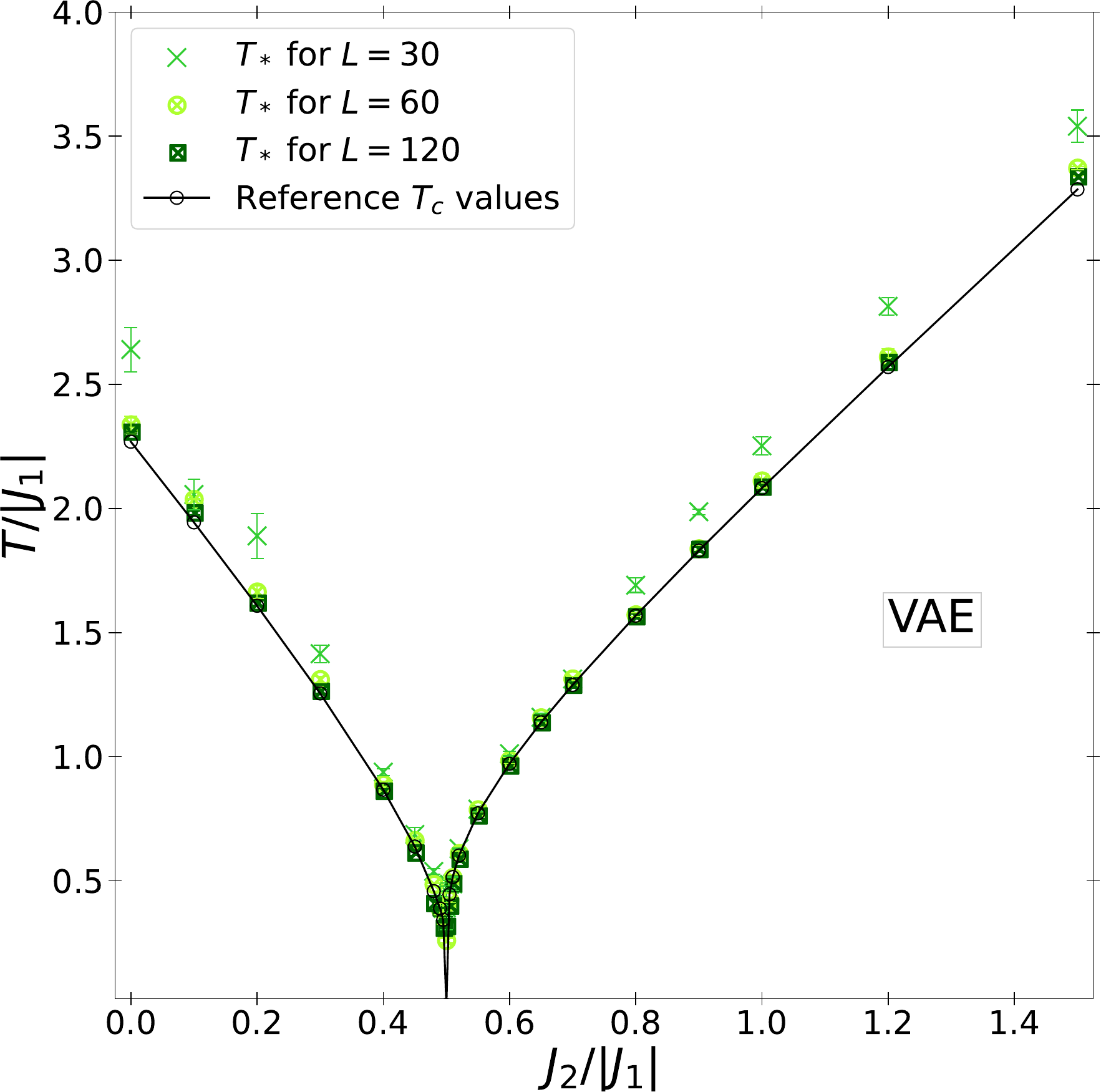}\hfill%
    \raisebox{0\columnwidth}{(b)}\includegraphics[width=0.95\columnwidth]{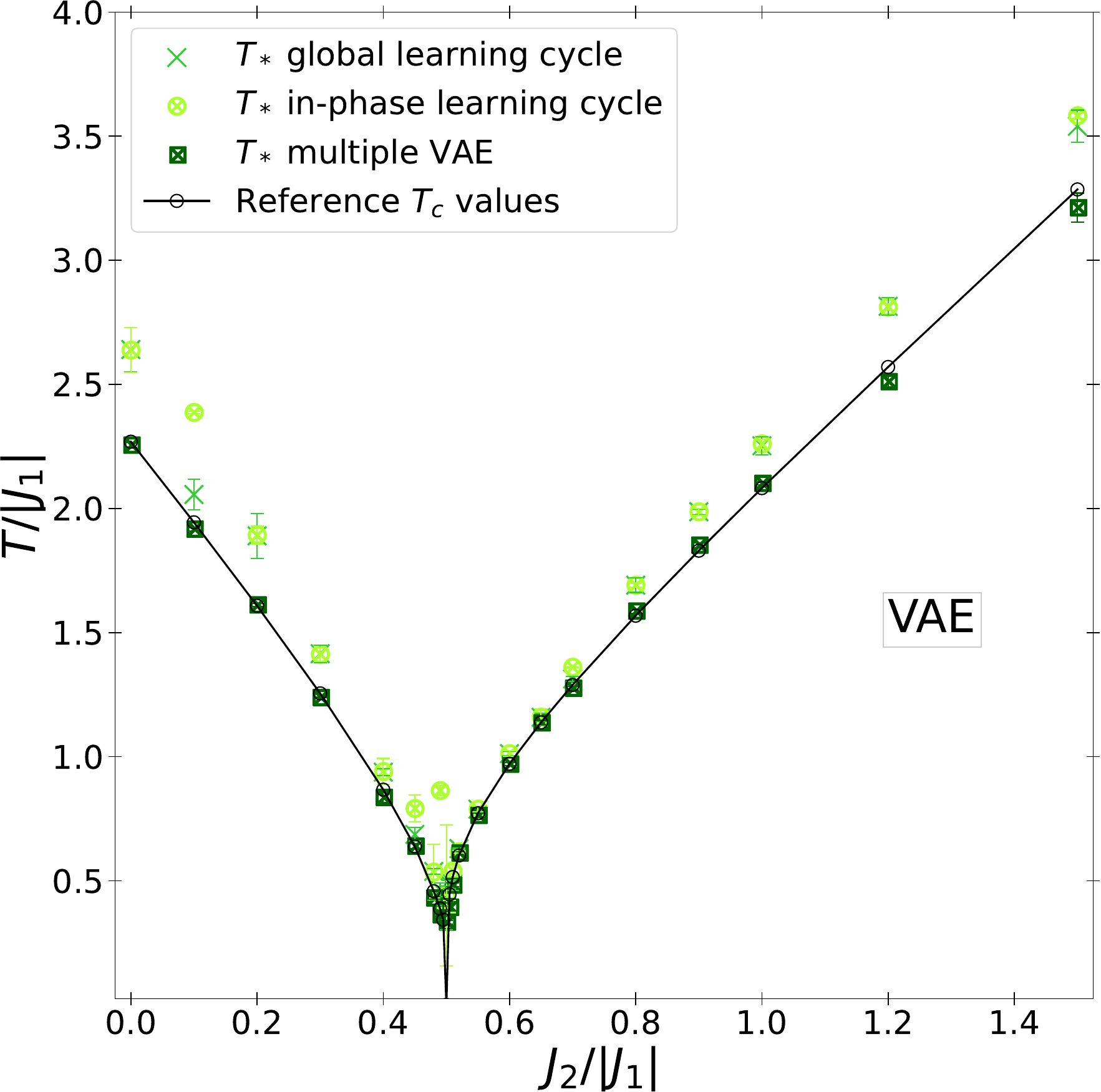}\\%\hfil%
    \raisebox{0\columnwidth}{(c)}\includegraphics[width=0.95\columnwidth]{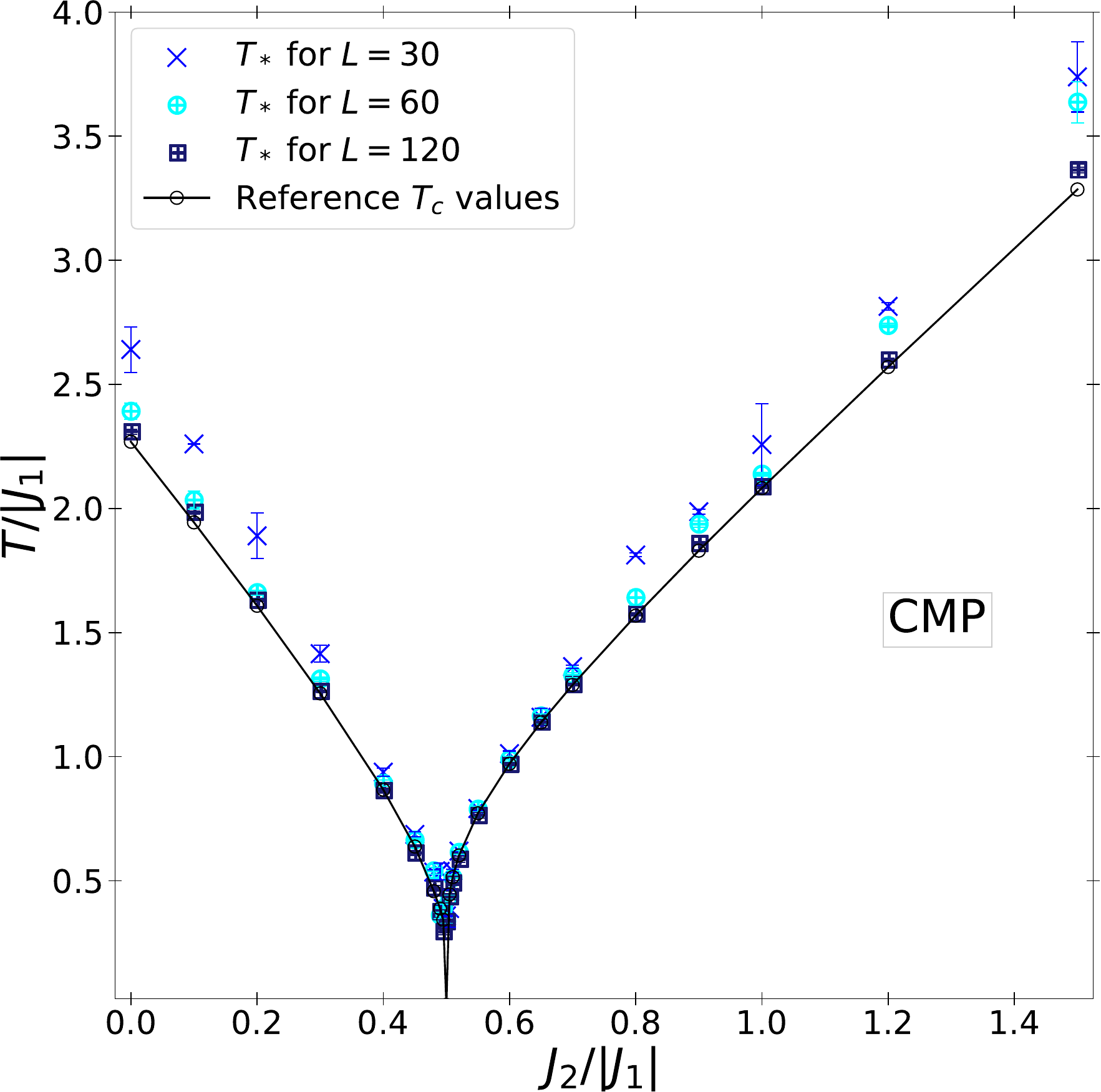}\hfill%
    \raisebox{0\columnwidth}{(d)}\includegraphics[width=0.95\columnwidth]{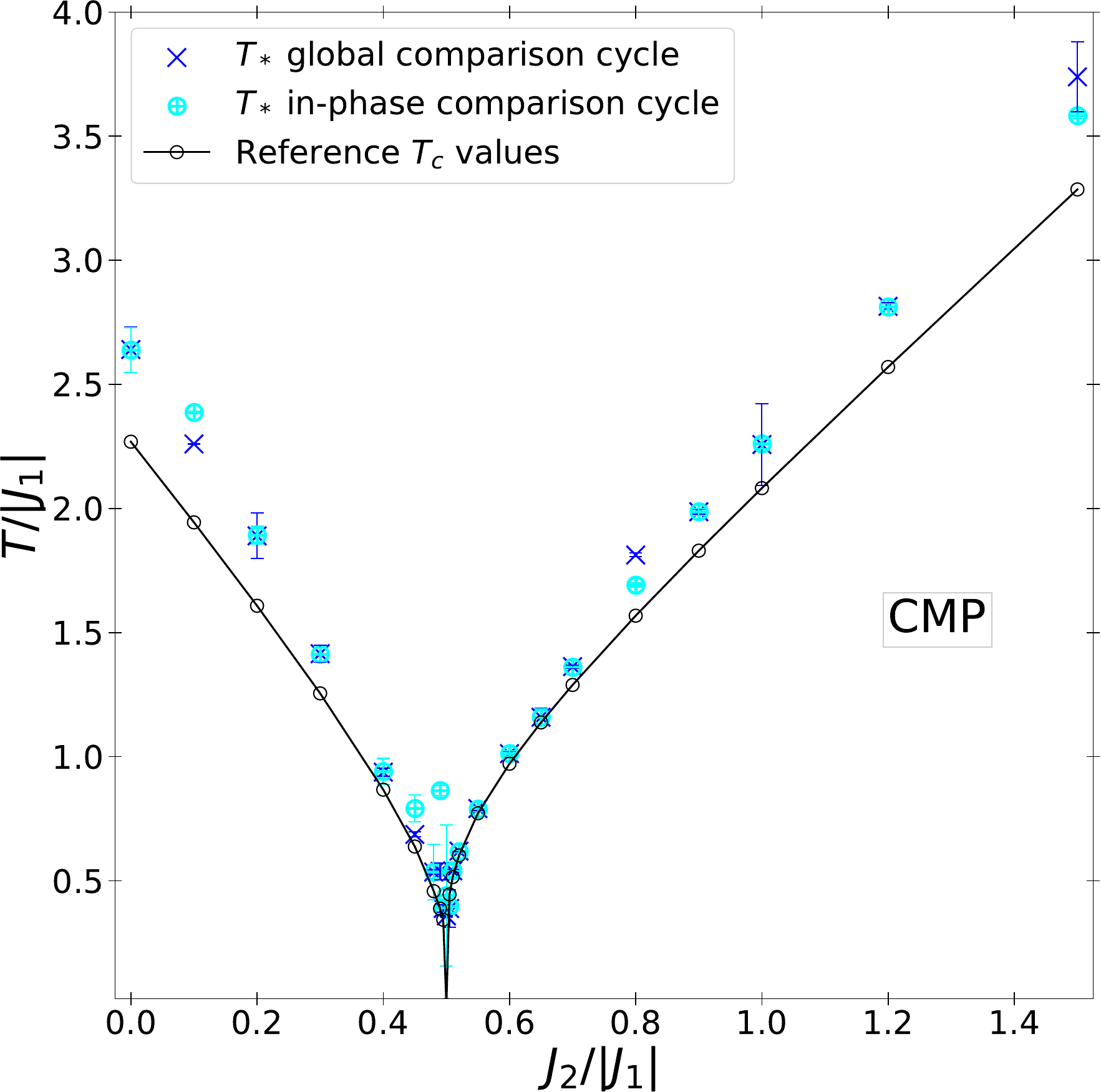}%\hfil%
    \end{center}
%  \vspace*{-1.5mm}
  \caption{%
  Comparison of predictions for phase boundaries as determined by the VAE (a+b) and CMP (c+d) approaches. 
  (a) Results  from VAE global training cycle (\ref{sec:singleVAE-global}) for different system sizes $L=30$ ($\times$, green), $60$ ($\bigotimes$, light green), and $120$ ($\boxtimes$, dark green). 
  (b) Results for
  $L=30$ from VAE global training cycle (\ref{sec:singleVAE-global}, $\times$, green), in-phase training cycle (\ref{sec:singleVAE-in-phase}, $\odot$, light green), and using multiple VAEs (\ref{sec:multipleVAE}, $*$, dark green).
  (c) Different $L$ for in-phase CMP (\ref{sec:ic-in-phase}) with $L=30$ ($\oplus$, blue), $60$ ($\bigoplus$, light blue), and $L=120$ ($\boxplus$, dark blue). 
  (d) Comparison of global CMP (blue (\text{$+$})) and in-phase CMP processes (light blue ($\odot$)) with $L=30$.
  In all panels, $\circ$ symbols connected by black lines denote the reference phase boundaries of Ref.~\cite{Kalz_2008}. 
  }
    \label{fig:results-VAE-IC-phase-boundaries}
\end{figure*}
%%%%%%%%%%%%%%%%%%%%%%%%%%%%%%%%%%%%%%%%%%%%%%%%%%%%%%%%%%%%%%%%%%%%%%

Figures \ref{fig:results-VAE-re} and \ref{fig:results-IC-re} indicate that different phases are distinguished well by different values of their reconstruction errors $\varepsilon$. Furthermore, we note that the sharpness of the difference in $\varepsilon$ becomes more pronounced when increasing $L$. This suggests that it is possible to obtain good estimates for the phase boundaries when computing the points $T$, $J_2$ at which these changes in $\varepsilon$ occur. In the following, we shall restrict ourselves to computing the temperature values $T_*$ at which the phases change, using the data as shown in Figs.\ \ref{fig:results-VAE-re} and \ref{fig:results-IC-re}, but for all $\mathcal{J}_2$.
We will estimate $T_*$  by computing the value of temperature at the maximum of the derivative,
\begin{equation}
     T_*(J_2) =
     \operatornamewithlimits{\rm argmax}\limits_T \left[ \frac{\partial{\mathcal{E}}(J_2)}{\partial T} \right].
     \label{eq:tstar}
\end{equation}
In order to reduce numerical fluctuations in the data, we use a simple cubic spline fit \cite{phillips1996theory}. For error estimation, the $\mathcal{E}(J_2)$ data is split into two frames with alternating $T$ values. We then calculate $T_{*1}$ and $T_{*2}$ in both frames and use half the absolute difference $|T_{*1}-T_{*2}|/2$ as error estimate
for $T_*$.

Figure \ref{fig:results-VAE-IC-phase-boundaries} shows the results of this analysis
for the $30\times30$, $60\times60$, and $120\times120$ lattices. Panel (a) shows that the $T_*$ values estimated by the derivative \eqref{eq:tstar} for the VAE-based $\varepsilon$ are indeed close to the known values for $T_c$ and the agreement gets better when increasing $L$ from $30$ to $120$,
i.e., the deviations seem to be mainly due to the VAE being applied to small lattices. Furthermore, the error estimates highlight that deviations from the reference data primarily arise from finite-size effects, but not exclusively.
The $L=30$ data in panel (a) shows that the extent of these deviations varies with $J_2$.
The $T_*$ estimates appear closer to the reference $T_c$ values in instances that are less noisy.
Figure \ref{fig:results-VAE-IC-phase-boundaries}(b) shows the
influence of the VAE training region on the $T_*$ predictions for the fixed size $L=30$.
One observes that, at fixed $L$, the high-$J_2$ in-phase training cycle reproduces the reference $T_c$ values best.

For the CMP approach, the differences between $T_*$ and $T_c$ are similar to the VAE: upon increasing $L$, the $T_*$ values become closer to their $T_c$ targets (Fig.~\ref{fig:results-VAE-IC-phase-boundaries}(c)), while the in-phase-based CMP method results in the best agreement of $T_*$ with $T_c$ (Fig.~\ref{fig:results-VAE-IC-phase-boundaries}(d)).
In Fig.~\ref{fig:results-VAE-IC-phase-boundaries}(c) one again observes not only better accuracy for the larger values of $L$, but also smaller statistical errors, like for the VAE in Fig.~\ref{fig:results-VAE-IC-phase-boundaries}(a).

%%%%%%%%%%%%%%%%%%%%%%%%%%%%%%%%%%%%%%%%%%%%%%%%%%%%%%%%%%%%%%%%%%%%%%
\section{Conclusions}
%%%%%%%%%%%%%%%%%%%%%%%%%%%%%%%%%%%%%%%%%%%%%%%%%%%%%%%%%%%%%%%%%%%%%%

We have explored two unbiased machine-``learning'' approaches to the detection of phase diagrams using the example of the $J_1$-$J_2$ Ising model on the square lattice.
We found that both the variational autoencoder (VAE) and a direct comparison of configurations (CMP) can successfully detect all three phases exhibited by this model, where the main factor limiting accuracy of the location of the phase boundaries appears to be the size of the lattices employed in these investigations, both via finite-size effects and a stronger influence of statistical noise for smaller systems.

Specifically, for the VAEs presented here,
we found in Sec.~\ref{sec:singleVAE} that, aiming for an unbiased reconstruction of the phases, a sequence of combinations of VAEs works best. Construction of this sequence still requires human input, so it is not yet a fully automatic ``machine-led'' process.

The use of VAEs to determine phases from just the spin configurations suggests that these themselves should contain sufficient information to identify phases.
Our second approach using just a simple comparison of configurations establishes that this indeed is possible.
In this approach, we replace the training phase of the VAE with a memory of spin configurations. We find that, for the relatively small system sizes considered here, both approaches give comparable accuracy. 

Both approaches yield, at least so far, limited accuracy when trying to determine the exact position of the transition points between phases.
It seems that here the VAE approach is somewhat better thanks to its built-in capability to interpolate, but there is still considerable room for improvement in both strategies. One might for example in the CMP approach replace the simple MSE with more sophisticated choices such as a zero normalized cross-correlation \cite{Papoulis1962} -- which of course could also be used as a quantity to
measure loss for the VAEs.

On reflection, both methods should be best used to determine the bulk of the phases and not so much to characterize the transition regions. As such, a more exploratory strategy suggests itself: knowing the critical temperature of the Ising model, $T_c(J_2=0)$, and the $T=0$ transition at $J_2=1/2$ in the $J_1$-$J_2$ model, one might want to find a starting point $(T,J_2)$ in the $T<T_c$ regime with (a) $J_2 \ll 1/2$ and (b) $J_2 \gg 1/2$. Then, e.g.\ for (a), one should explore locally around that starting point, with either VAE or CMP, and find other $(T', J'_2)$ such that $\varepsilon(T,J_2)\sim \varepsilon(T',J'_2)$. This would explore the phase close to the starting $(T,J_2)$. Then repeat the same for region (b). Clearly, when $\varepsilon(T,J_2)\nsim \varepsilon(T',J'_2)$, one comes close to the phase boundaries. Such a strategy, when using the CMP, would only need to store the starting configuration at $(T,J_2)$ with very little memory consumption. One could even include the newly found configurations belonging to the same phase when comparing with further configurations.

In conclusion, our investigations indicate that in previous work based on the reconstruction error \cite{Corte2021,Acevedo_2021_2} a neuronal network is not fundamentally required, but that the essential idea behind uncovering the structure of the phase diagram, without manually defining an order parameter for each phase, is to actually \emph{look} at the configurations, and this process can be automated even without resorting to a neuronal network.
Implementation of a direct comparison of configurations is straightforward,
could still be optimized beyond the present implementation,
and avoids possible complications inherent to VAEs such as the need to ensure proper training of the neuronal network.

Let us mention similar caveats that some of the present authors have recently found in the context of percolation. Not only does a neural-network approach fail to correctly reproduce the sample-to-sample fluctuations of the correlation length $\xi$ in a supervised-learning context \cite{Bayo_2022,Bayo_2023}, but instead of learning the physics of a \emph{global} spanning cluster in classification of percolating versus non-percolating configurations, the network seems to rather learn how to guess this property via the proxy of the density of occupied sites in the system.

These findings call for further investigations to understand the real value of neuronal networks, in particular those designed for image-recognition tasks, when these are applied to the study of phase transitions.

%%%%%%%%%%%%%%%%%%%%%%%%%%%%%%%%%%%%%%%%%%%%%%%%%%%%%%%%%%%%%%%%%%%%%%
\begin{acknowledgments}
%\ack
We thank D.\ Bayo (Warwick) for helpful discussions. 
RAR gratefully acknowledges funding by the CY Initiative of Excellence (grant ``Investissements d'Avenir'' ANR-16-IDEX-0008) and hospitality during various short-term stays at CY Advanced Studies. 
RAR also gratefully acknowledges support from the University of Warwick Research Technology Platform (RTP Scientific Computing). 
UK research data statement: No new data was generated.
\end{acknowledgments}
%%%%%%%%%%%%%%%%%%%%%%%%%%%%%%%%%%%%%%%%%%%%%%%%%%%%%%%%%%%%%%%%%%%%%%

%%%%%%%%%%%%%%%%%%%%%%%%%%%%%%%%%%%%%%%%%%%%%%%%%%%%%%%%%%%%%%%%%%%%%%
\appendix

%%%%%%%%%%%%%%%%%%%%%%%%%%%%%%%%%%%%%%%%%%%%%%%%%%%%%%%%%%%%%%%%%%%%%%
\section{The sign of $J_1$}
\label{app:A}
%%%%%%%%%%%%%%%%%%%%%%%%%%%%%%%%%%%%%%%%%%%%%%%%%%%%%%%%%%%%%%%%%%%%%%

Here we show that changing the sign of $J_1$ in the Hamiltonian (\ref{eq:hamiltonian}) yields equivalent physics.
It is convenient to split the site index $i$ into two integer coordinates $x$, $y=1$, $\ldots$, $L$. Then consider the following transformation of spin variables
\begin{equation}
s'_{x,y} = (-1)^{x+y} \, s_{x,y}\, .
\end{equation}
When rewritten in terms of these new variables, the Hamiltonian (\ref{eq:hamiltonian}) 
becomes
\begin{eqnarray}
H_{J_1J_2} &=& +J_1 \,\sum_{x,y=1}^L  s'_{x,y} \, \left(s'_{x+1,y} + s'_{x,y+1}\right)
\nonumber \\
&& + J_2 \,\sum_{x,y=1}^L s'_{x,y} \,  \left(s'_{x+1,y+1} + s'_{x+1,y-1}\right) \,
. \qquad
\label{eq:hamiltonianP}
\end{eqnarray}
Clearly, the sign of $J_1$ has changed while the one of $J_2$ has remained unchanged.
This implies that energy-related observables are independent of the sign of $J_1$, including the phase diagram.
However, the precise nature of the configuration is changed; for example, the ferromagnetic state
in the conventions of the main text is mapped to the antiferromagnetic one in the primed variables.

We note in passing that in order to implement periodic boundary conditions, one needs to identify
the coordinate $L+1$ with $1$ and $0$ with $L$.

\nocite{*}
\bibliography{ML}% Produces the bibliography via BibTeX.

\end{document}